\documentclass[a4paper,11pt]{article}
\usepackage[utf8x]{inputenc}
\usepackage{cite}
\usepackage{url}
\usepackage{color}
\usepackage{graphicx}
\usepackage{fullpage}
\usepackage{amsmath,amssymb}
\usepackage{authblk}

\def\beq{\begin{equation}}
\def\beqn{\begin{eqnarray}}
\def\eeq{\end{equation}}
\def\eeqn{\end{eqnarray}}

\def\PDF#1#2{\Gamma_{\!#1/#2}}
\def\hPDF#1#2{\hat{\Gamma}_{\!#1/#2}}
\def\oPDF#1#2{\overline{\Gamma}_{\!#1/#2}}
\def\beamPDF#1#2#3{\phi_{\!#1/#2}^{(#3)}}
\def\hbeamPDF#1#2#3{\hat{\phi}_{\!#1/#2}^{(#3)}}
\def\obeamPDF#1#2#3{\overline{\phi}_{\!#1/#2}^{(#3)}}

\newcommand\sss{\scriptscriptstyle}

\newcommand\as{\alpha_{\sss S}}
\newcommand\aem{\alpha}

\newcommand\mcatnlo{{\sc\small MC@NLO}}

\newcommand\aNLO{{\sc\small MadGraph5\_aMC@NLO}}
\newcommand\aNLOs{{\sc\small MG5\_aMC}}
\newcommand\UFO{{\sc\small UFO}}
\newcommand\MadGraph{{\sc\small MadGraph}}
\newcommand\MadGraphf{{\sc\small MadGraph5}}
\newcommand\epem{e^+e^-}
\newcommand{\lp}{e^+}
\newcommand{\lm}{e^-}
\newcommand{\lpm}{e^{\pm}}
\newcommand{\lmp}{e^{\mp}}
\newcommand{\zp}{z_+}
\newcommand{\zm}{z_-}
\newcommand{\zpm}{z_{\pm}}
\newcommand{\yp}{y_+}
\newcommand{\ym}{y_-}
\newcommand{\ypm}{y_{\pm}}
\newcommand{\xp}{x_+}
\newcommand{\xm}{x_-}
\newcommand{\xpm}{x_{\pm}}

\newcommand{\xpb}{p}

\newcommand\hsig{\hat{\sigma}}

\newcommand\MSb{\overline{\rm MS}}
\newcommand\bkappa{\bar{\kappa}}

\title{Lepton collisions in MadGraph5\_aMC@NLO}
\author[1]{Stefano Frixione\thanks{Stefano.Frixione@cern.ch}}
\author[2]{Olivier Mattelaer\thanks{olivier.mattelaer@uclouvain.be}}
\author[3]{Marco Zaro\thanks{marco.zaro@mi.infn.it}}
\author[4]{Xiaoran Zhao\thanks{xiaoran.zhao@uniroma3.it}}

\affil[1]{\small INFN Sezione di Genova, Via Dodecaneso 33, 16146, 
Genova, Italy}
\affil[2]{\small Centre for Cosmology, Particle Physics and Phenomenology 
(CP3), Universit\'e Catholique de Louvain, Chemin du Cyclotron, 
1348 Louvain la Neuve, Belgium}
\affil[3]{\small TIFLab, Dipartimento di Fisica, Universit\`a degli
Studi di Milano and INFN, Sezione di Milano,\\ Via Celoria 16, 
20133 Milano, Italy}
\affil[4]{\small Dipartimento di Matematica e Fisica, Universit{\`a} di 
Roma Tre and INFN, sezione di Roma Tre, I-00146 Rome, Italy }

\begin{document}

\maketitle

\begin{abstract}
\noindent
MadGraph5\_aMC@NLO is a software package that allows one to simulate 
processes of arbitrary complexity, at both the leading and the next-to-leading
order perturbative accuracy, with or without matching and multi-jet merging 
to parton showers. It has been designed for, and so far primarily employed
in the context of, hadronic collisions. In this note, we document the
implementation of a few technical features that are necessary to extend
its scope to realistic $e^+e^-$ collider environments. We limit ourselves 
to discussing the unpolarized-beam case, but we point out that the treatment
of polarized beams is conceptually identical, and that the structure we
set up can easily be extended to carry out simulations at $\mu^+\mu^-$
colliders.
\end{abstract}

\section{Introduction\label{sec:intro}}
In view of the fact that \aNLO~\cite{Alwall:2014hca} (referred to as
\aNLOs\ henceforth) has not been extensively employed so far in the context
of $\epem$ collisions, we give here the briefest of summary of its present
scope and characteristics. \aNLOs\ constructs automatically the short-distance
cross section for an arbitrary process, given in input by the user together
with the Lagrangian of the theory (in the form of a \UFO\ 
model~\cite{Degrande:2011ua}) in which the computation is performed;
such a cross section is subsequently integrated and, where appropriate,
unweighted events associated with it are stored as Les Houches 
event (LHE henceforth) files~\cite{Alwall:2006yp,Butterworth:2010ym}.
The form of the cross section depends on the type of simulation one
wishes to perform, which is in turn determined by the combination of its 
perturbative accuracy (leading order (LO) or next-to-leading order (NLO)),
of whether it is matched or not to a parton shower Monte Carlo (MC) and,
in the case where such a matching is involved, whether multi-jet merging
is also required. The interested reader can find full details in
ref.~\cite{Alwall:2014hca}.

If the underlying accuracy is an LO one, the features just described
are qualitatively similar to those of the \MadGraph\ family. However,
it is important to keep in mind that \aNLOs\ now vastly exceeds the 
capabilities of the latter codes; in particular, it supersedes their
most recent ones, \MadGraphf~\cite{Alwall:2011uj}, and must be used
in its place. Conversely, at the NLO \aNLOs\ collects the features 
developed in the course of several works -- the 
automation~\cite{Frederix:2009yq} of the FKS~\cite{Frixione:1995ms,
Frixione:1997np} subtraction, and of the parton-shower matching (and merging)
according to the \mcatnlo~\cite{Frixione:2002ik} (FxFx~\cite{Frederix:2012ps})
formalism. In keeping with the general philosophy that underpins the
code, the type of perturbative series is driven by the given \UFO\ model;
furthermore, \aNLOs\ is capable of handling simultaneously the expansion 
in two coupling constants, most notably those of QCD and of the electroweak 
Standard Model (EW SM). The latter one is expected to be increasingly 
relevant as the LHC enters a high-precision phase, and it is obviously
particularly important for $\epem$ computations; its (technically involved) 
automation has been presented in ref.~\cite{Frederix:2018nkq}.

What has been said thus far broadly summarises the status of \aNLOs\
as a widely-used tool for hadronic-collision simulations. Remarkably,
at the level of short-distance cross sections {\em no changes} are
necessary for it to deal with the case of $\epem$ collisions, regardless
of the perturbative accuracy (LO or NLO) and of the underlying theoretical
model (e.g.~QCD and/or EW). However, while $\epem$ short-distance cross
sections (at variance with their hadronic counterparts) can give one a 
very good idea of the corresponding physics, strictly speaking they are 
non-physical as soon as one takes into account {\em EW-induced corrections}, 
owing to their lacking the description of two phenomena, namely: {\em a)} 
beamstrahlung effects\footnote{More precisely, beam-beam interactions,
of which beamstrahlung is one of the consequences, and typically the dominant
one on the energy spectra of the incoming particles. Because of this 
dominance, in a slightly improper manner we shall identify the two 
concepts in this note.}; {\em b)} multiple emissions 
of low-energy/low-angle photons and leptons from the particles that are 
about to enter the hard process -- such multiple emissions will be 
collectively called initial-state radiation (ISR) effects henceforth. 
We point out that only ISR, at variance with beamstrahlung, can be 
unambiguously classified as perturbative. More details will be given 
in sect.~\ref{sec:xsec}. 

In summary, the public version of \aNLOs\ could compute, before the present
work, both LO- and NLO-accurate $\epem$ cross sections, but without 
including beamstrahlung and ISR effects. The goal of this note is to 
document the implementation of such effects in the code in view of its 
next release\footnote{That will be v3.2.0 of \aNLOs. We point out that 
an independent implementation of ISR effects in \aNLOs\ has been 
presented in refs.~\cite{Chen:2017ipx,Li:2018qnh}. The latter have a
limited scope, and are superseded by the present work.}, that will be 
limited to leading-logarithmic (LL) and LO results\footnote{This strictly 
applies to the code implementation: the formulae presented in this note 
are general, and encompass all cases. The code will soon be upgraded to 
included higher-order effects in $\aem$.} in $\aem$, while no restriction 
will be imposed on the accuracy in $\as$.

\section{$\epem$ cross sections\label{sec:xsec}}
Both beamstrahlung and ISR are due to initial-state emissions.
Conventionally, all perturbative and collider-independent effects
are associated with the latter, whereas the former accounts for
collider-specific beam-dynamics phenomena. The evaluation of ISR
effects necessarily involves an analysis to all orders in perturbation
theory, and thus entails some degree of approximation. The approach
we follow in \aNLOs\ is based on parametrising ISR by means of a
factorisation formula, quite similar to its QCD counterpart. Conversely,
the understanding of beamstrahlung is rather heuristic, and typically
stems from an MC simulation of beam-beam interactions, which is strictly
collider dependent. Beamstrahlung and ISR are independent from each other;
however, while the prediction of a physical $\epem$ cross section must
include ISR effect, beamstrahlung ones might or might not be required.
Therefore, in \aNLOs\ two options are considered: either ISR-only, or
ISR combined with beamstrahlung.

Following ref.~\cite{Frixione:2019lga}, we write the differential
cross section for the production of a system of particles $X$ in
$\epem$ collisions:
\beq
e^+(P_{\lp})\,+e^-(P_{\lm})\;\longrightarrow\; X
\label{eeX}
\eeq
as follows:
\beq
d\Sigma_{\epem}(P_{\lp},P_{\lm})=\sum_{kl}\int d\yp d\ym\,
{\cal B}_{kl}(\yp,\ym)\,d\sigma_{kl}(\yp P_{\lp},\ym P_{\lm})\,,
\label{beamstr}
\eeq
with
\beq
k\;\in\;\{\lp,\gamma\}\,,\;\;\;\;\;\;\;\;
l\;\in\;\{\lm,\gamma\}\,,
\label{klrange}
\eeq
and where the sum over polarisation states is understood. In
eq.~(\ref{beamstr}), the function ${\cal B}_{kl}(\yp,\ym)$ parametrises
the emergence of the pair of particles $k$ and $l$ due to the dynamics
of the incoming $\epem$ beams. Such two particles carry a fraction (equal 
to $\yp$ and $\ym$, respectively) of the longitudinal momentum of the incoming
beams. In turn, particles $k$ and $l$ undergo a hard collision, whose
cross section is denoted by $d\sigma_{kl}$ and which includes ISR effects.
The ranges of their possible identities in eq.~(\ref{klrange}) stem from 
considering only the dominant contributions to the vast majority of production 
processes; they can be extended if need be.
As was anticipated, we parametrise $d\sigma_{kl}$ by writing the cross
section as follows:
\beqn
d\sigma_{kl}(\xpb_k,\xpb_l)&=&\sum_{ij}\int d\zp d\zm\,
\PDF{i}{k}(\zp,\mu^2,m^2)\,\PDF{j}{l}(\zm,\mu^2,m^2)
\nonumber\\*&&\phantom{\sum_{ij}\int}\times
d\hsig_{ij}(\zp\xpb_k,\zm\xpb_l,\mu^2,m^2)\,.
\label{factth0}
\eeqn
In eq.~(\ref{factth0}), the $\PDF{i}{k}$ symbol denotes the
PDF of {\em parton} $i$ in {\em particle} $k$; the former carries a 
fraction $\zpm$ of the longitudinal momentum of the latter. Such a quantity
is thus conceptually similar to its QCD counterpart, and indeed it obeys
the same DGLAP~\cite{Gribov:1972ri,Lipatov:1974qm,Altarelli:1977zs,
Dokshitzer:1977sg} evolution equations\footnote{We limit ourselves to
only including QED effects in our ISR parametrisation. If need be,
purely-weak ones can be accounted for with the same formalism.\label{ft:QED}}:
\beq
\frac{\partial \PDF{i}{k}}{\partial\log\mu^2}=
\frac{\aem}{2\pi}P_{ij}\otimes\PDF{j}{k}\,.
\label{APeq2}
\eeq
Importantly, the PDFs of interest here can be fully computed in a 
perturbative manner, at variance with the QCD ones. The quantity denoted 
by $d\hsig_{ij}$ in eq.~(\ref{factth0}) is the (IR-subtracted) partonic
short-distance cross section, associated with the following hard process:
\beq
i(\zp\xpb_k)\,+j(\zm\xpb_l)\;\longrightarrow\; X\,.
\label{eeXpart}
\eeq
Finally, by $m$ and $\mu$ we have denoted the electron mass and a scale
of the order of the hardness of the process. In \aNLOs, the latter can
be set by the user. While for full generality in eq.~(\ref{factth0})
we have included an electron-mass dependence in the partonic cross
section, in practice in \aNLOs\ such a dependence is always ignored:
all electron-mass effects are included by means of the PDFs (analogous
final-state effects would have to be included by means of fragmentation
functions).

In summary, the physics content of eq.~(\ref{beamstr}), and our
naming conventions for it, can be compactly written as follows:
\beq
{\rm beams}\;\stackrel{{\rm beamstrahlung}}{\longrightarrow}\;
{\rm particles}\;\stackrel{{\rm ISR}}{\longrightarrow}\;
{\rm partons}\,.
\eeq
Consistently with this, we call:
\begin{itemize}
\item $d\Sigma_{\epem}$: collider-level cross section.
\item $d\sigma_{kl}$: particle-level cross section.
\item $d\hsig_{ij}$: (subtracted) parton-level cross section.
\end{itemize}
Note that, at variance with the QCD case, the identities of a particle and 
of a parton might coincide (in particular, the PDF of an electron/positron
in an electron/positron will generally give the dominant contributions
at an $\epem$ collider).

According to eq.~(\ref{klrange}), one or two photons might emerge from
the beamstrahlung and initiate the hard process. In practice, in the 
current version of \aNLOs\ we limit ourselves to considering the case 
\mbox{$(k,l)=(e^+,e^-)$} -- in other words, the identities of the 
particles coincide with those of the beams whence they emerge;
formally:
\beq
{\cal B}_{kl}(\yp,\ym)=\delta_{k\lp}\delta_{l\lm}{\cal B}_{\epem}(\yp,\ym)\,.
\label{Bee}
\eeq
Conversely, the identities of the partons are dictated by the combination 
of the nature of the system $X$, of the required perturbative
accuracy, and of the (formal) perturbative expansions of the partonic
cross sections and of the PDFs. At the LO in $\aem$ (i.e.~the only option 
we make publicly available as a result of the present work) this is trivial, 
as it only involves electrons/positrons. We shall specify in a future note 
the strategy we shall pursue when dealing with higher-order QED corrections.

The current public version of \aNLOs\ is able to compute partonic 
cross sections. In this note, we document the implementation of the 
beamstrahlung and ISR effects according to eqs.~(\ref{beamstr})
and~(\ref{factth0}), respectively, within the assumptions discussed
above.

\section{ISR\label{sec:ISR}}
As was anticipated in sect.~\ref{sec:xsec}, we currently ignore the
particle-level cross sections which are not initiated by an $\epem$ pair.
Therefore, we implement eq.~(\ref{factth0}) with \mbox{$(k,l)=(e^+,e^-)$} 
and can thus limit ourselves to employing electron PDFs (see 
footnote~\ref{ft:QED}). For such PDFs, in \aNLOs\ the choice 
is given to adopt either the LO+LL results of refs.~\cite{Skrzypek:1990qs,
Skrzypek:1992vk,Cacciari:1992pz} or (in future versions) the NLO+NLL ones of 
refs.~\cite{Bertone:2019hks,Frixione:2021wzh}; more details on these PDFs 
can be found in those papers.

The structure of eq.~(\ref{factth0}) is the same as that of the factorisation
theorems of QCD, upon which the simulations of hadronic collisions in
\aNLOs\ is based. Unfortunately, the corresponding implementation cannot
be used as is in the context of $\epem$ collisions, owing to the vastly
different behaviour of hadron and electron PDFs -- while the former peak
at small $z$'s, the latter peak at $z\to 1$. Furthermore, while in hadronic
collisions the very-small-$z$ integration region is cut-off by the requirement
that there be a minimal invariant mass produced by the hard process, in
$\epem$ collisions the region around $z=1$ is never cut-off; it actually 
gives the dominant contribution to the cross section by far\footnote{The
presence of an $s$-channel resonance with mass smaller than the collider
energy can partly act as an effective cut-off; we shall discuss this
case in the following.}. This is due to the fact that the PDF of an 
electron/positron in an electron/positron reads as follows in the 
$z\to 1$ region:
\beq
\PDF{\lpm}{\lpm}(z)\;\;\stackrel{z\to 1}{\longrightarrow}\;\;
\frac{\hPDF{\lpm}{\lpm}(z)}{(1-z)^{1-\beta}}
\label{eePDFasy}
\eeq
both at the LO+LL and at the NLO+NLL 
(see refs.~\cite{Bertone:2019hks,Frixione:2021wzh}). 
The function $\hPDF{\lpm}{\lpm}(z)$ is at most logarithmically-divergent
at $z\to 1$, and $\beta$ is a parameter whose precise definition (and possible 
dependence on the hard scale of the process) depends on the perturbative order,
but which is always numerically small (at the typical $\epem$ colliders
considered nowadays, $\beta\simeq 0.05$). Thus, eq.~(\ref{eePDFasy})
exhibits a very pronounced integrable divergence at $z=1$, which is
reason for the dominance mentioned before. Conversely, the PDFs
$\PDF{\gamma}{\lpm}$ and $\PDF{\lmp}{\lpm}$ that potentially appear 
in an NLO computation are at most logarithmically divergent at $z\to 1$.
Therefore, in the following we shall discuss only eq.~(\ref{eePDFasy}) 
which, in addition to being the only one relevant to LO computations,
is also a worst-case scenario.

In order to integrate eq.~(\ref{factth0}) with PDFs that behave as
in eq.~(\ref{eePDFasy}), \aNLOs\ performs $\epem$-specific changes
of variables. One starts by writing the PDFs as follows:
\beq
\PDF{\lpm}{\lpm}(z)=
\frac{\oPDF{\lpm}{\lpm}(z)}{(1-z)^{\gamma}}\,,
\label{eePDFint}
\eeq
with $\gamma$ a free constant parameter such that:
\beq
1-\beta\le\gamma <1\,.
\label{garange}
\eeq
Note that, owing to the smallness of $\beta$, $\gamma$ must always be
chosen rather close to one.
This implies that the functions $\oPDF{\lpm}{\lpm}(z)$ implicitly
defined in eq.~(\ref{eePDFint}) are at most logarithmically divergent
at $z=1$ when \mbox{$\gamma=1-\beta$}, and vanish there otherwise.
Then, one introduces the variables:
\beq
t_\pm=\left(\frac{1-\zpm}{1-z_{0\pm}}\right)^{1-\gamma}
\;\;\;\;\Longrightarrow\;\;\;\;
\frac{d\zpm}{dt_\pm}=
\frac{1-z_{0\pm}}{1-\gamma}\,t_\pm^{\gamma/(1-\gamma)}=
\frac{\left(1-z_{0\pm}\right)^{1-\gamma}}{1-\gamma}\,
\left(1-\zpm\right)^\gamma\,,
\label{tpmdef}
\eeq
with $z_{0\pm}\ge 0$ the lower limit of the integration range in $\zpm$.
Thus, for each of the two incoming-particle legs:
\beqn
d\zpm\,\PDF{\lpm}{\lpm}(\zpm)\,f_\pm(\zpm)&\equiv&
d\zpm\,\frac{\oPDF{\lpm}{\lpm}(z)}{(1-z)^{\gamma}}\,f_\pm(\zpm)
\nonumber
\\*&=&
dt_\pm\,
\oPDF{\lpm}{\lpm}(\zpm(t_\pm))\,f_\pm(\zpm(t_\pm))\,
\frac{\left(1-z_{0\pm}\right)^{1-\gamma}}{1-\gamma}\,,
\label{chvar}
\eeqn
where $f_\pm$ are arbitrary regular functions. Further changes
of variables $t_\pm\to u_\pm$ with:
\beq
t_\pm=u_\pm^k
\;\;\;\;\Longrightarrow\;\;\;\;
\frac{dt_\pm}{du_\pm}=ku_\pm^{k-1}
\eeq
and $k\gtrsim 2$ help the convergence in the case of NLO+NLL PDFs which,
in the $\MSb$ scheme, feature \mbox{$\log^p(1-z)$} terms (whereas they do 
not in the scheme introduced in ref.~\cite{Frixione:2021wzh}, at least
when the initial scale for the evolution is set equal to the electron mass).

The r.h.s.~of eq.~(\ref{chvar}) is now well-behaved at $z\to 1$, and can
easily be integrated numerically\footnote{Roughly speaking, the dominant
$z\to 1$ integral contributions are turned into the large
\mbox{$1/(1-\gamma)$} factors.}. However, for this to happen it is
crucial that $\oPDF{\lpm}{\lpm}(z)$ be well behaved for all $z$'s,
in particularly for those arbitrarily close to one\footnote{It is 
instructive to point out that the integration of eq.~(\ref{factth0})
with $d\hsig_{ij}=1$ and the constraint \mbox{$\zp\zm\le 1-10^{-\kappa}$}
is within $0.1$\% of its asymptotic $\kappa=\infty$ value for
$\kappa={\cal O}(50)$. This figure is so large that even the simple
computation of $1-z$, let alone anything more complicated than that, 
turns out to be beyond the accuracy of standard double-precision
computations.\label{ft:prec}}. For this reason, the factor
\beq
\frac{(1-z)^{\gamma}}{(1-z)^{1-\beta}}=
\left(1-z\right)^{\gamma+\beta-1}
\label{omzfct}
\eeq
which is present in the $z\to 1$ expression of $\oPDF{\lpm}{\lpm}(z)$
must always be computed as in the r.h.s.~of eq.~(\ref{omzfct}), and
not as in its l.h.s.; in turn, this requires the analytical knowledge 
of the large-$z$ behaviour of the PDFs. These observations have been
taken into account when the results for the (N)LO+(N)LL PDFs of 
refs.~\cite{Bertone:2019hks,Frixione:2021wzh}) have been implemented in 
numerical routines;
those return $\oPDF{\lpm}{\lpm}(z)$ (and not $\PDF{\lpm}{\lpm}(z)$),
given $\gamma$, $z$, and $1-z$; the latter quantity is computed in
\aNLOs\ not as such, but directly as the $t$'s of eq.~(\ref{tpmdef}),
for the reasons explained in footnote~\ref{ft:prec}.

Given eqs.~(\ref{garange}) and~(\ref{omzfct}), it appears that the
ideal choice for $\gamma$ is to set it equal to $1-\beta$. Unfortunately,
this is impossible beyond LO, in view of the fact that $\beta$ 
generally becomes scale-dependent. Since the scale may be chosen on
an event-by-event basis (i.e.~as a function of the kinematical
configuration) in keeping with the standard assumptions in \aNLOs,
the value of $\beta$ is known only after having generated to partonic
four-momenta, that in turn require the variables $\zpm$ to have been
already computed. Thus, by setting \mbox{$\gamma=1-\beta$} one is
led to a circular argument.

We conclude this part by giving a few more details on the integration
ranges of the $\zpm$ variables. The partonic cross section on the r.h.s.~of
eq.~(\ref{factth0}) implicitly implements a cut-off:
\beq
\tau\ge\tau_0\,,
\;\;\;\;\;\;\;\;
\tau=\zp\zm\,,
\;\;\;\;\;\;\;\;
\tau_0=\frac{M_{\rm low}^2}{\left(\xpb_k+\xpb_l\right)^2}\,,
\label{taus}
\eeq
with $M_{\rm low}^2$ the smallest possible invariant mass of the
system $X$ produced in the hard collision (such a quantity is strictly
larger than zero). We can thus set:
\beq
z_{0+}=\tau_0\,,
\;\;\;\;\;\;\;\;
z_{0-}=\frac{\tau_0}{\zp}\,.
\label{z0s}
\eeq
The asymmetric choice of eq.~(\ref{z0s}) is not ideal for analytical
differential calculations. In particular, if $\zp\to\tau_0$ then
$\zm$ will always be in a neighbourhood of one; thus, {\em small}
$\zp$ values induce a sensitivity to the large-$z$ (divergent)
behaviour of the PDFs. This fact does not turn out to be problematic
in numerical computations, provided that one follows the strategy
outlined so far. In particular, if $\zp\to\tau_0$ the integration 
range in $\zm$ becomes vanishingly small, which is accounted for
(in a numerically-stable way, when expressing $\zp$ in terms of $t_+$) 
by the factor \mbox{$(1-z_{0-})^{1-\gamma}$} in eq.~(\ref{tpmdef}).

\subsection{The case of $s$-channel resonances}
The integration procedure described above does not perform well in
the presence of $s$-channel resonances that emerge from the incoming
particles. In order to present the strategy employed by \aNLOs\ in
such a case, we start by observing that the problem can always be
treated as if there were only one resonance. In fact, in the case of 
multiple resonances \aNLOs\ is able to disentagle them by means of a 
multi-channeling strategy (see refs.~\cite{Maltoni:2002qb,Frederix:2009yq} 
for the LO and NLO approaches, respectively), and to treat them separately.

We denote by $M$ and $\Gamma$ the mass and the width, respectively,
of the resonance of interest, and by:
\beq
s=\left(P_{\lp}+P_{\lm}\right)^2
\eeq
the collider c.m.~energy squared\footnote{This coincides with the
particle c.m.~energy squared when beamstrahlung effects are
ignored.}. In view of eq.~(\ref{taus}), we also introduce the
auxiliary quantities:
\beq
\tau_M=\frac{M^2}{s}\,,\;\;\;\;\;\;\;\;
\tau_\Gamma=\frac{\Gamma^2}{s}\,.
\eeq
We assume that $M<\sqrt{s}$ (since otherwise the resonance does not
entail any special treatment), and therefore $\tau_M$ belongs to the
integration range of $\zp\zm$. Thus, in terms of $\tau$ the integrand
of eq.~(\ref{factth0}) has two peaks, at $\tau=\tau_M$ (stemming from
the $s$-channel resonant propagator, which gives rised to a Breit-Wigner
function in the partonic cross sections) and at $\tau=1$ (stemming from
the PDFs). These are competing effects, and in order to integrate both of
them efficienctly, we proceed as follows. We first introduce the quantities:
\beqn
f_{\rm res}&=&\frac{1}{(\tau-\tau_M)^2+\tau_M\tau_\Gamma}\,,
\label{fres}
\\
f_{\rm nr}&=&\frac{1}{(1-\tau)^{1-2\beta}}\,.
\label{fnr}
\eeqn
Equation~(\ref{fres}) is nothing but a rescaled Breit-Wigner function,
whereas the form of eq.~(\ref{fnr}) is motivated by the observation
that the presence of a resonance induces an effective luminosity
strongly peaked at $\tau=1$:
\beq
{\cal L}(\tau)=\int d\zp d\zm \delta\left(\tau-\zp\zm\right)
\PDF{\lp}{\lp}(\zp)\PDF{\lm}{\lm}(\zm)
\stackrel{\tau\to 1}{\longrightarrow}\frac{1}{(1-\tau)^{1-2\beta}}\,,
\eeq
where use of eq.~(\ref{eePDFasy}) has been made. The current integration
channel is then split into two terms, by multiplying the partonic cross
section by a factor equal to one, written as follows:
\beq
1=F_{\rm res}+F_{\rm nr}\,,
\eeq
with:
\beqn
F_{\rm res}&=&\frac{f_{\rm res}}{f_{\rm res}+f_{\rm nr}}\,,
\\
F_{\rm nr}&=&\frac{f_{\rm nr}}{f_{\rm res}+f_{\rm nr}}\,.
\eeqn
In other words, the r.h.s.~of eq.~(\ref{factth0}) is written as the sum 
of two terms, each of which identical to the r.h.s.~of eq.~(\ref{factth0}) 
bar for the insertion of either $F_{\rm res}$ or $F_{\rm nr}$ in the
integrand. Such terms are then integrated separately\footnote{In practice,
this is done on the fly. We generate a random number $r$; if $r<r_M$  
($r>r_M$) we integrate the $F_{\rm res}$ ($F_{\rm nr}$) term. The value
of $r_M$ can be set either in an adaptive manner, or fixed (e.g.~$r_M=1/2$), 
the two choices being conceptually equivalent to each other.}. 
For the one relevant to $F_{\rm nr}$ we proceed as was 
explained in sect.~\ref{sec:ISR}, while for that relevant to $F_{\rm res}$ 
we first change integration variables \mbox{$(\zp,\zm)\to (\tau,z_i)$}, with 
either $z_i=\zp$ or $z_i=\zm$ (chosen at random, and exploiting the fact
that at the LO+LL the situation of the two beams is symmetric). 
Next, we flatten the integration in $\tau$ by generating it by means of 
the inverse of the integral function of eq.~(\ref{fres}), and that in 
$z_i$ by using the change of variables of eq.~(\ref{tpmdef}).

\section{Beamstrahlung\label{sec:beamstr}}
We now consider the case of a non-trivial beamstrahlung function
in eq.~(\ref{beamstr}); it is in fact possible to treat such a case
in the same formal way as the trivial one (i.e.~as if beamstrahlung
were not present), since the latter corresponds to setting:
\beq
{\cal B}_{kl}(\yp,\ym)=\delta_{k\lp}\delta_{l\lm}
\delta(1-\yp)\,\delta(1-\ym)\,.
\label{Btriv}
\eeq
As we shall see the similarity between the trivial and non-trivial
cases is even stronger, since non-trivial beamstrahlung functions feature
combinations of regular functions and of distributions.

As was said above, information on the beamstrahlung function is obtained
by means of MC simulations: codes able to perform such simulations include
CAIN~\cite{Chen:1994jt}, GuineaPig~\cite{Schulte:1997nga}, and
GuineaPig++~\cite{Schulte:2007zz}. The format in which the results
of the MC simulations are stored in order for the cross-section
integrators to use them is not standard, and one should employ the
structure best suited to one own's needs. For example, in 
CIRCE1~\cite{Ohl:1996fi} a beta function is adopted to fit the
one-dimensional energy spectrum for each beam, thereby ignoring correlations 
between the two beams. Such an approach has later been 
extended~\cite{Sailer:2009zz} to a two-dimensional fit that allows one
to account for beam correlations. Conversely, in CIRCE2~\cite{Kilian:2007gr}
a grid-based strategy is adopted instead.

A peculiar characteristics of eq.~(\ref{beamstr}), as opposed to its
particle-level counterpart eq.~(\ref{factth0}), is that the beamstrahlung
functions ${\cal B}_{kl}(\yp,\ym)$ are not separable in their arguments
$\ypm$. We must therefore use the fact that the following representation
holds~\cite{Sailer:2009zz}:
\beq
{\cal B}_{kl}(\yp,\ym)\approx\sum_{n=1}^N\,
b^{(\lp)}_{n,kl}(\yp)\,b^{(\lm)}_{n,kl}(\ym)\,,
\label{Bexp}
\eeq
for suitable basis functions $b^{(\lpm)}_{n,kl}(y)$. Equation~(\ref{Bexp})
becomes an identity for $N\to\infty$, but in practice one can obtain
a highly accurate representation of ${\cal B}_{kl}$ even with a relatively
small value of $N$. In particular, for the only case
(\mbox{$(k,l)=(e^+,e^-)$}, see eq.~(\ref{Bee})) implemented in \aNLOs\
v3.2.0, we have found the following form to give satisfactory
results for both linear and circular colliders\footnote{We assume
$0\le\ypm\le 1$, and therefore ignore beam-spread effects.}:
\beqn
{\cal B}_{\epem}(\yp,\ym)&=&
\hat{f}_{11}\,\delta(1-\yp)\,\delta(1-\ym)
\nonumber\\*&+&
(1-\yp)^{\kappa_+}f_{01}(\yp)\,\delta(1-\ym)
\nonumber\\*&+&
\delta(1-\ym)\,(1-\ym)^{\kappa_-}f_{10}(\ym)
\nonumber\\*&+&
(1-\yp)^{\kappa_+}f_{00+}(\yp)\,(1-\ym)^{\kappa_-}f_{00-}(\ym)\,,
\label{Bee2}
\eeqn
with~\cite{Yokoya:1989jb}
\beq
\kappa_+=\kappa_-=-2/3\,,
\label{kvals}
\eeq
and where the functions $f_\alpha(y)$ are regular, and generally depend
on the collider type. By equating the r.h.s.'s of eqs.~(\ref{Bexp})
(with $N=4$) and~(\ref{Bee2}) one can solve for the functions
$b^{(\lpm)}_{n,\epem}(\ypm)$ that appear in the former; we need not do it 
explicitly here. The complete determination of ${\cal B}_{\epem}(\yp,\ym)$
is achieved by choosing suitable forms for the functions $f_\alpha(y)$,
that depend on a finite number of parameters. After having done that,
beam dynamics is simulated by means of GuineaPig~\cite{Schulte:1997nga}
with high-statistics runs (with up to $100$M events). The results of 
these are first separated according to whether the condition(s)
$\ypm=1$ is(are) met, then when either $\yp\ne 1$ and/or $\ym\ne 1$ are
histogrammed in the $\yp$ and/or $\ym$ variables, and subsequently
fit separately with the last three lines of eq.~(\ref{Bee2}), in order 
to determine the free parameters of the chosen functional forms. Conversely,
the $\hat{f}_{11}$ parameter is simply determined by the counting of
events with $\yp=1$ and $\ym=1$. It is clear that this operation, while
time-consuming, must be done only once per collider type. We have
considered several examples; these are summarised in table~\ref{tab:collpars},
which also reports the values of the beam-dynamics parameters used in
the GuineaPig simulations.
\begin{table}[t!]
\begin{center}
\begin{tabular}{c|c|c|c|c|c}
& Name~\cite{Schulte:1997nga} & FCC-ee & CEPC & ILC & CLIC \\
\hline
$E_{beam}$[GeV] & energy & 120,\,182.5 & 120 & 125,\,250 & 1500 \\
$N$[$10^{10}$] & particles & 15,\,27 & 15 & 2 & 0.37 \\
$\sigma_E$[$10^{-3}$] & espread & 1.65,\,2.0 & 1.0 & 
[$e^{-}$:1.9,1.2][$e^{+}$:1.5,0.7] & 3.5 \\
$\sigma_x$[nm] & sigma\_x & 14000,\,38000 & 21000 & 520,\,470 & 40 \\
$\sigma_y$[nm] & sigma\_y & 40,\,70 & 70 & 8,\,6 & 1 \\
$\sigma_z$[$\mu$m] & sigma\_z & 5300,\,3800 & 3300 & 300 & 44 \\
$\beta_x$[mm] & beta\_x & 300,\,1000 & 360 & 13,\,22 & 6.9 \\
$\beta_y$[mm] & beta\_y & 1,\,1.6 & 1.5 & 0.41,\,0.48 & 0.068\\
$\theta$[rad] & angle\_x & 0.015 & 0.0165 & 0.007 & 0.010 \\
\end{tabular}
\end{center}
\caption{\label{tab:collpars}$\epem$ collider configurations, with the 
respective parameters used in the GuineaPig simulations: $E_{beam}$ is
the beam energy, $N$ the number of particles per bunch, $\sigma_E$ the
beam-energy spread,  $\sigma_x, \sigma_y$ the beam sizes, $\sigma_z$ the
bunch length, $\beta_x, \beta_y$ the amplitude functions,
and $\theta$ the crossing angle.}
\end{table}

\begin{figure}[th!]
\begin{center}
\includegraphics[width=0.45\textwidth]{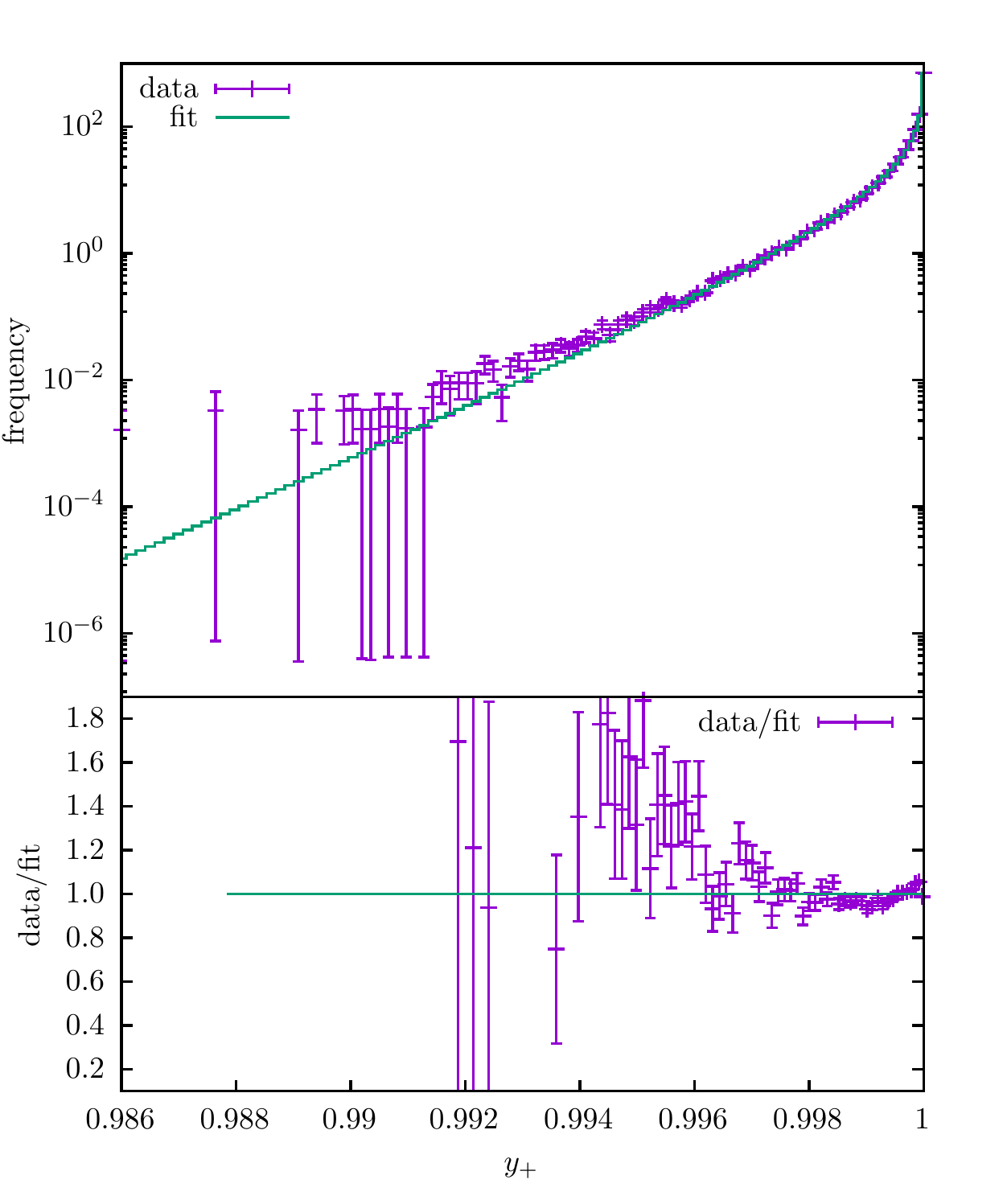}\hspace{5mm}
\includegraphics[width=0.45\textwidth]{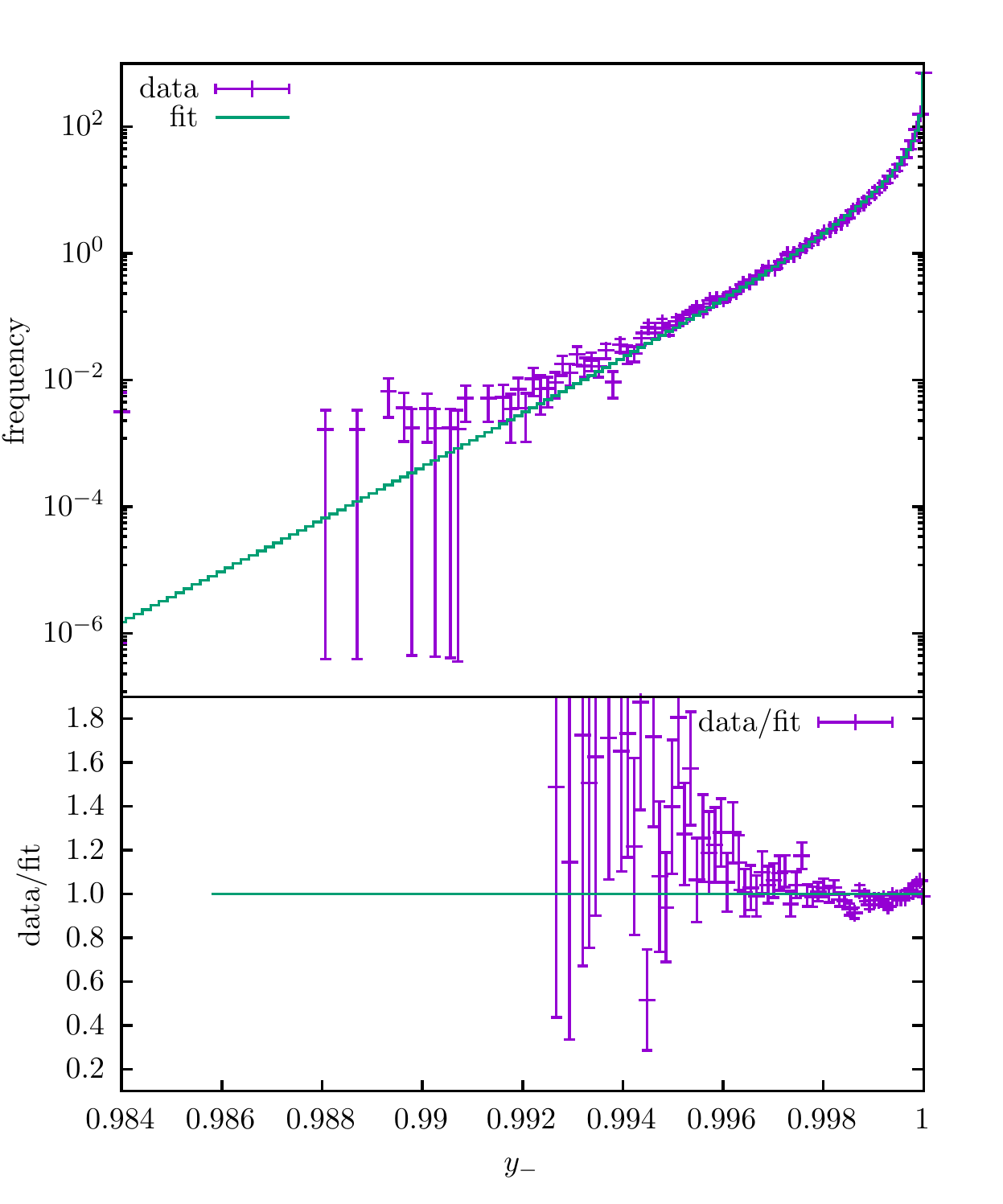}
\caption{\label{fig:f2dfccee240}
Fits to the $\yp$ (left panel) and $\ym$ (right panel) components of
the fourth line of eq.~(\ref{Bee2}), for FCC-ee240.
}
\end{center}
\end{figure}

\begin{figure}[th!]
\begin{center}
\includegraphics[width=0.45\textwidth]{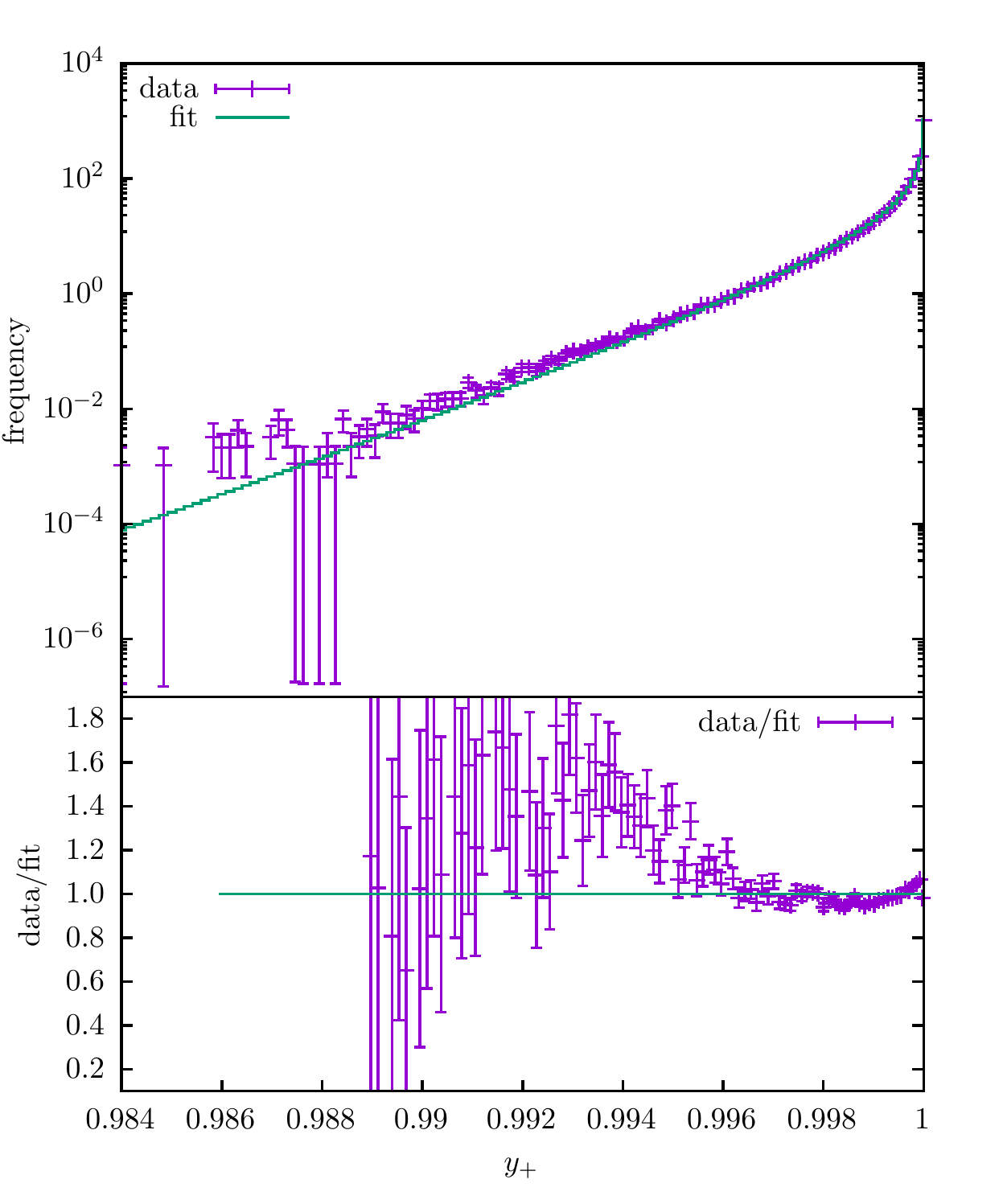}\hspace{5mm}
\includegraphics[width=0.45\textwidth]{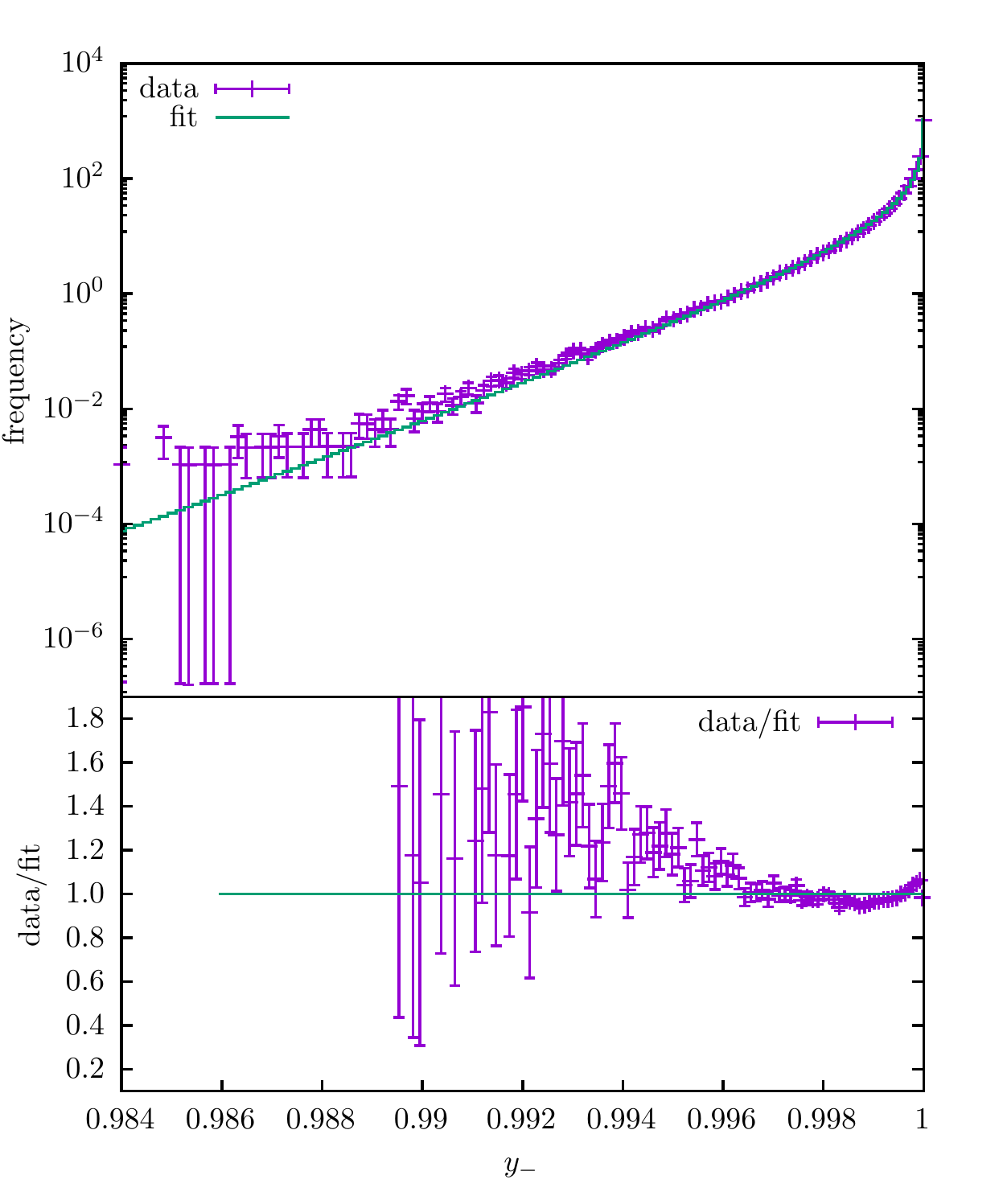}
\caption{\label{fig:f2dfccee365}
As in fig.~\ref{fig:f2dfccee240}, for FCC-ee365.
}
\end{center}
\end{figure}

\begin{figure}[th!]
\begin{center}
\includegraphics[width=0.45\textwidth]{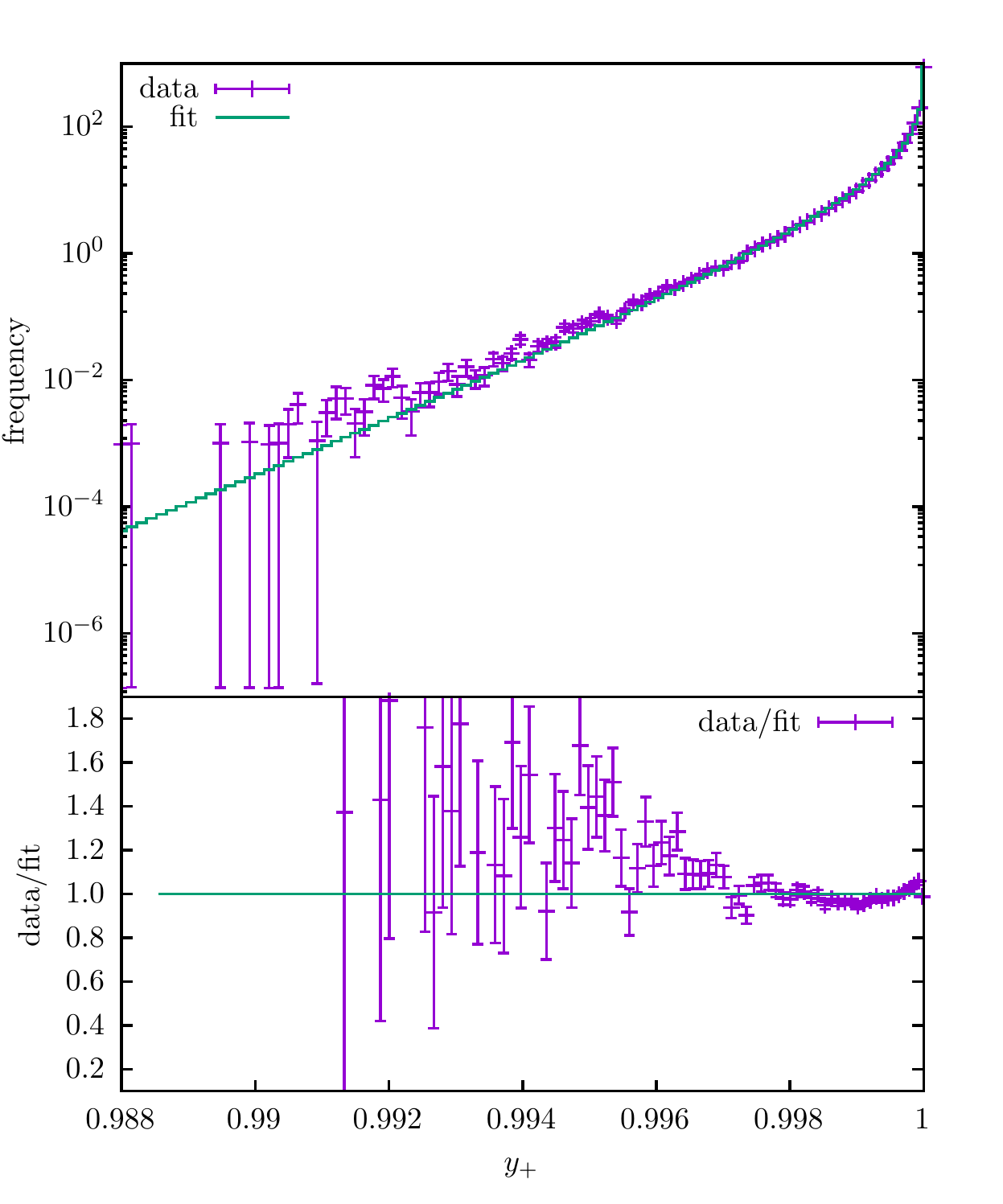}\hspace{5mm}
\includegraphics[width=0.45\textwidth]{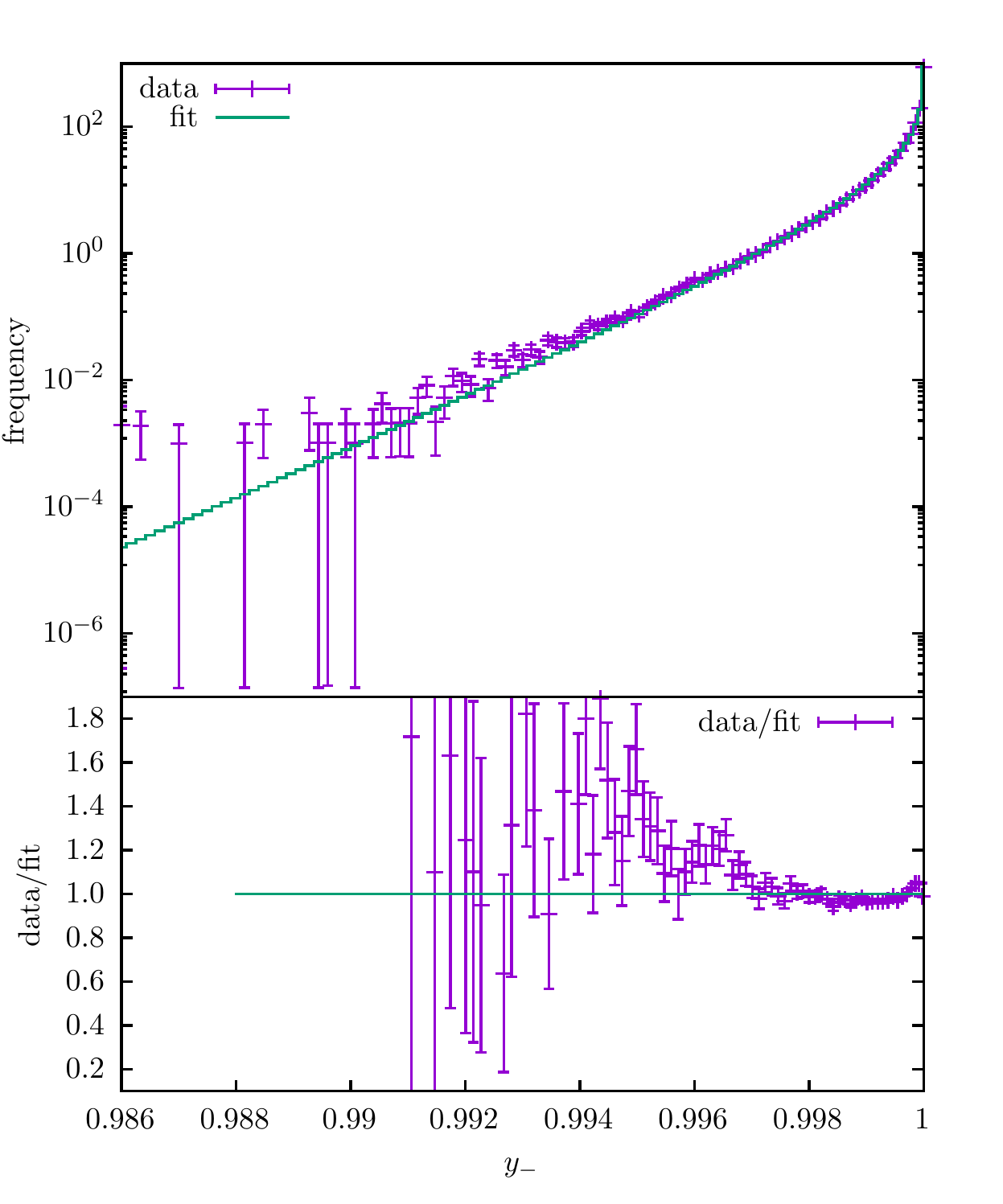}
\caption{\label{fig:f2dcepc240}
As in fig.~\ref{fig:f2dfccee240}, for CEPC240.
}
\end{center}
\end{figure}

In order to be more specific, let us first consider the case of
circular colliders. By means of the function
\beq
f(y;p,q)=e^{p(1-y)}\,e^{q\sqrt{1-y}}
\label{ffuncirc}
\eeq
we set:
\beqn
f_{01}(y)&=&\hat{f}_{01}\,f(y;p_{01},q_{01})\,,
\label{f01fun}
\\
f_{10}(y)&=&\hat{f}_{10}\,f(y;p_{10},q_{10})\,,
\label{f10fun}
\\
f_{00+}(y)&=&\hat{f}_{00+}\,f(y;p_{00+},q_{00+})\,,
\label{f00pfun}
\\
f_{00-}(y)&=&\hat{f}_{00-}\,f(y;p_{00-},q_{00-})\,.
\label{f00mfun}
\eeqn
The values of the parameters $\hat{f}_i$, $p_i$, and $q_i$ that 
result from the fits to the GuineaPig simulations as described above
are reported in tables~\ref{tab:fit-fccee240}, \ref{tab:fit-fccee365},
and~\ref{tab:fit-cepc240} for the circular-collider configurations
of table~\ref{tab:collpars}, namely FCC-ee with $E_{beam}=120$~GeV
(denoted by FCC-ee240), FCC-ee with $E_{beam}=182.5$~GeV
(denoted by FCC-ee365), and CEPC with $E_{beam}=120$~GeV
(denoted by CEPC240), respectively. The corresponding comparisons 
between the fitted functional forms and the simulation data are 
presented in appendix~\ref{sec:fitres}, in figs.~\ref{fig:f1dfccee240}
and~\ref{fig:f2dfccee240}, \ref{fig:f1dfccee365} and~\ref{fig:f2dfccee365}, 
and~\ref{fig:f1dcepc240} and~\ref{fig:f2dcepc240}.
\begin{table}[!htb]
\begin{center}
    \begin{tabular}{|c|c|c|c|c|}
        \hline
        $i$ & $\hat{f}_i$ & $p_i$ & $q_i$ & Integral\\
        \hline
        $11$ & 0.8698 &  &  & \\
        \hline
        $01$ & 0.2863 & $-901.2$ & $-19.30$ & 0.06234\\
        \hline
        $10$ & 0.2853 & $-916.5$ & $-18.70$ & 0.06230 \\
        \hline
        $00+$ & 0.3308 & $-899.2$ & $-18.03$ & 0.07308\\
        $00-$ & 0.3303 & $-918.7$ & $-17.48$ & 0.07306\\
        \hline
        Sum &  &  &  & 0.9998 \\
        \hline
    \end{tabular}
\end{center}
\caption{\label{tab:fit-fccee240}Fit results for FCC-ee240.}
\end{table}
\begin{table}[!htb]
\begin{center}
    \begin{tabular}{|c|c|c|c|c|}
        \hline
        $i$ & $\hat{f}_i$ & $p_i$ & $q_i$ & Integral\\
        \hline
        $11$ & 0.7883 &  &  &  \\
        \hline
        $01$ & 0.4056 & $-714.3$ & $-14.64$ & 0.09856\\
        \hline
        $10$ & 0.4059 & $-714.5$ & $-14.69$ & 0.09856 \\
        \hline
        $00+$ & 0.4772 & $-714.7$ & $-12.77$ & 0.1188\\
        $00-$ & 0.4780 & $-717.1$ & $-12.80$ & 0.1189\\
        \hline
        Sum &  &  &  & 0.9996 \\
        \hline
    \end{tabular}
\end{center}
    \caption{\label{tab:fit-fccee365}Fit results for FCC-ee365.}
\end{table}
\begin{table}[!htb]
\begin{center}
\begin{tabular}{|c|c|c|c|c|c|c|}
    \hline
        $i$ & $\hat{f}_i$ & $p_i$ & $q_i$ & Integral\\
      \hline
    $11$ & 0.8351 &  &  & \\
    \hline
    $01$ & 0.3513 & $-910.9$ & $-17.28$ & 0.07806\\
    \hline
    $10$ & 0.3515 & $-914.2$ & $-17.27$ & 0.07803\\
    \hline
    $00+$ & 0.4075 & $-917.9$ & $-15.44$ & 0.09230\\
    $00-$ & 0.4080 & $-918.8$ & $-15.45$ & 0.09239\\
    \hline
    Sum &  &  &  & 0.9997 \\
    \hline
\end{tabular}
\end{center}
\caption{\label{tab:fit-cepc240}Fit results for CEPC240.}
\end{table}

\begin{figure}[th!]
\begin{center}
\includegraphics[width=0.45\textwidth]{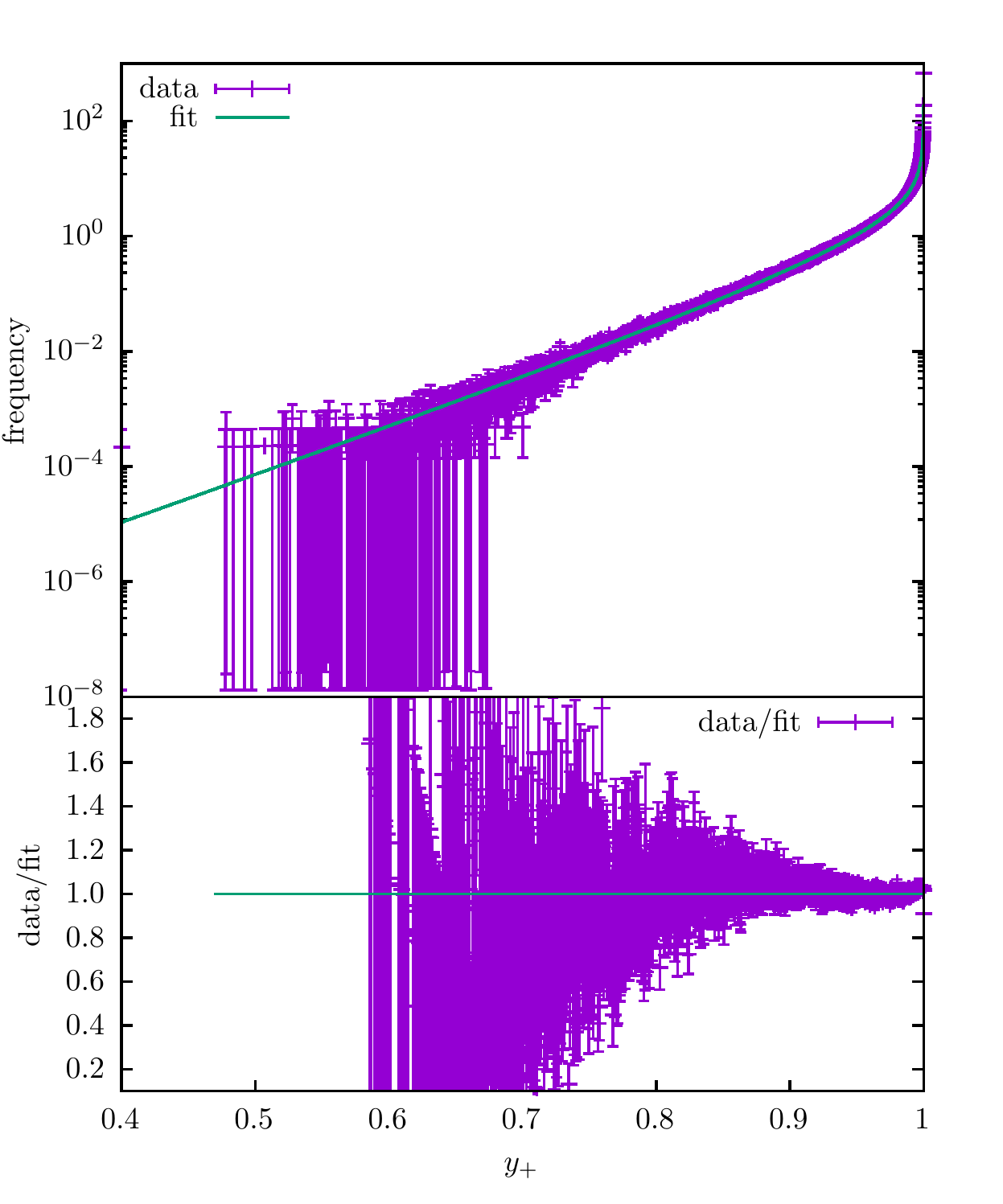}\hspace{5mm}
\includegraphics[width=0.45\textwidth]{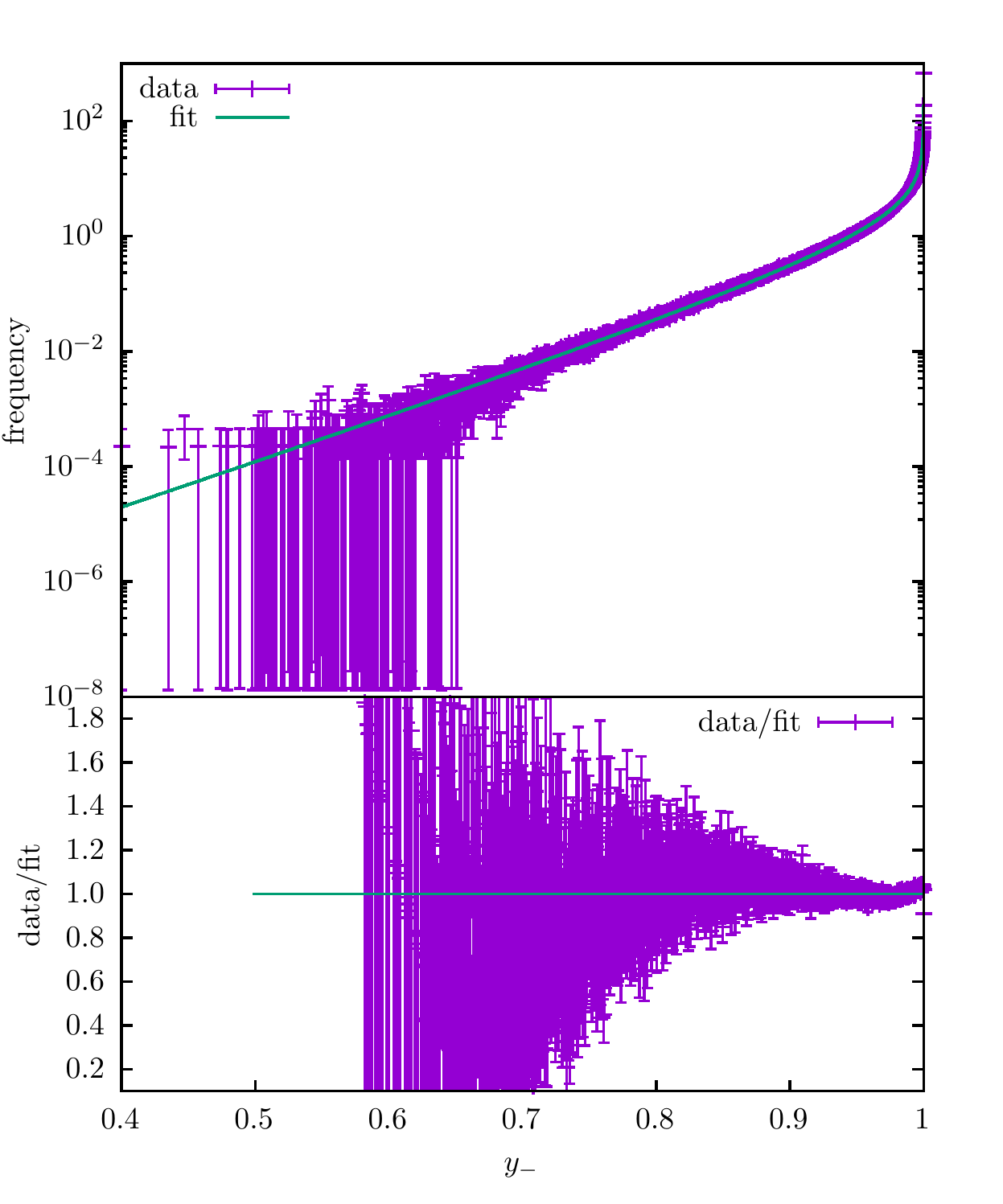}
\caption{\label{fig:f2dilc250}
As in fig.~\ref{fig:f2dfccee240}, for ILC250.
}
\end{center}
\end{figure}

\begin{figure}[th!]
\begin{center}
\includegraphics[width=0.45\textwidth]{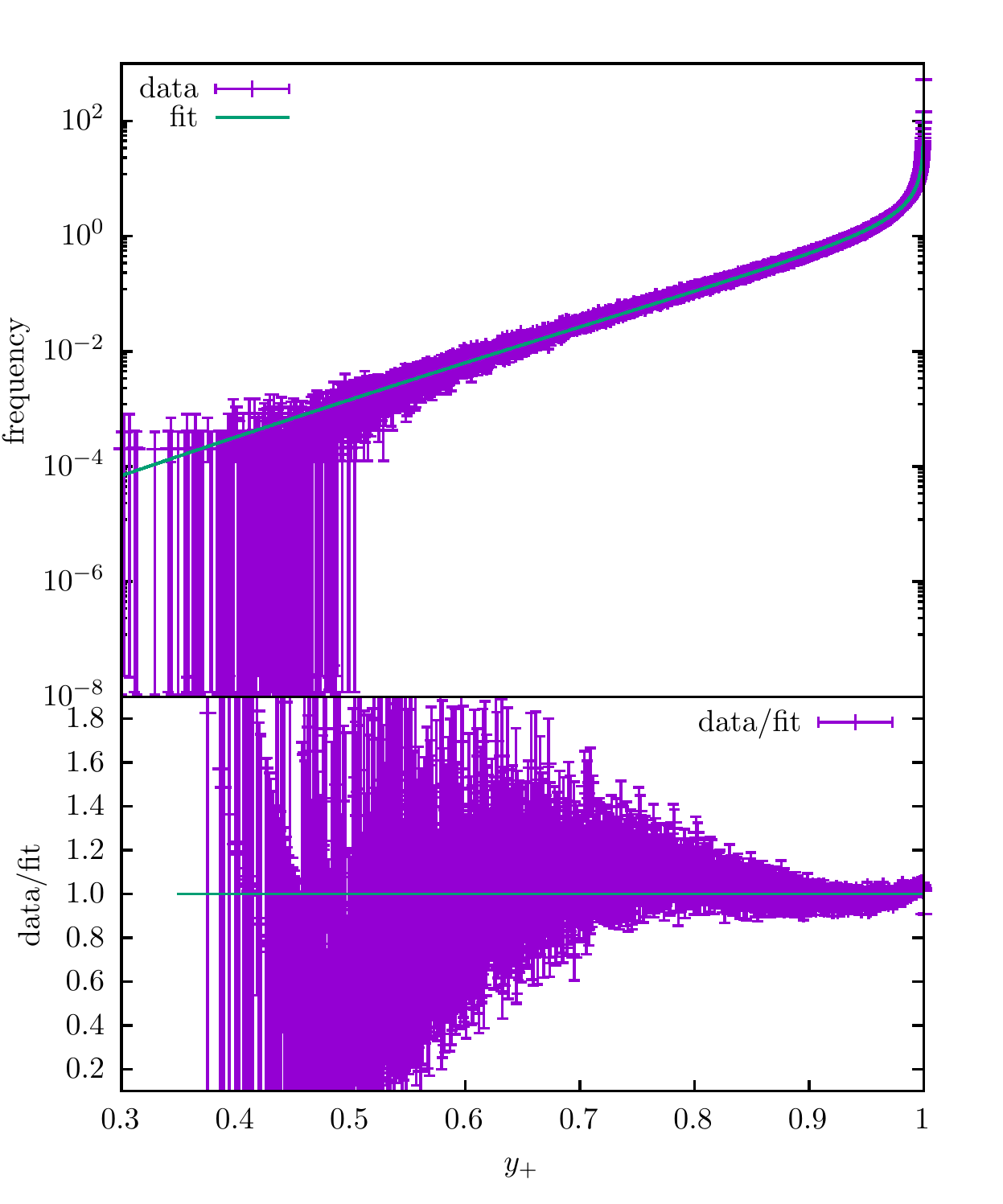}\hspace{5mm}
\includegraphics[width=0.45\textwidth]{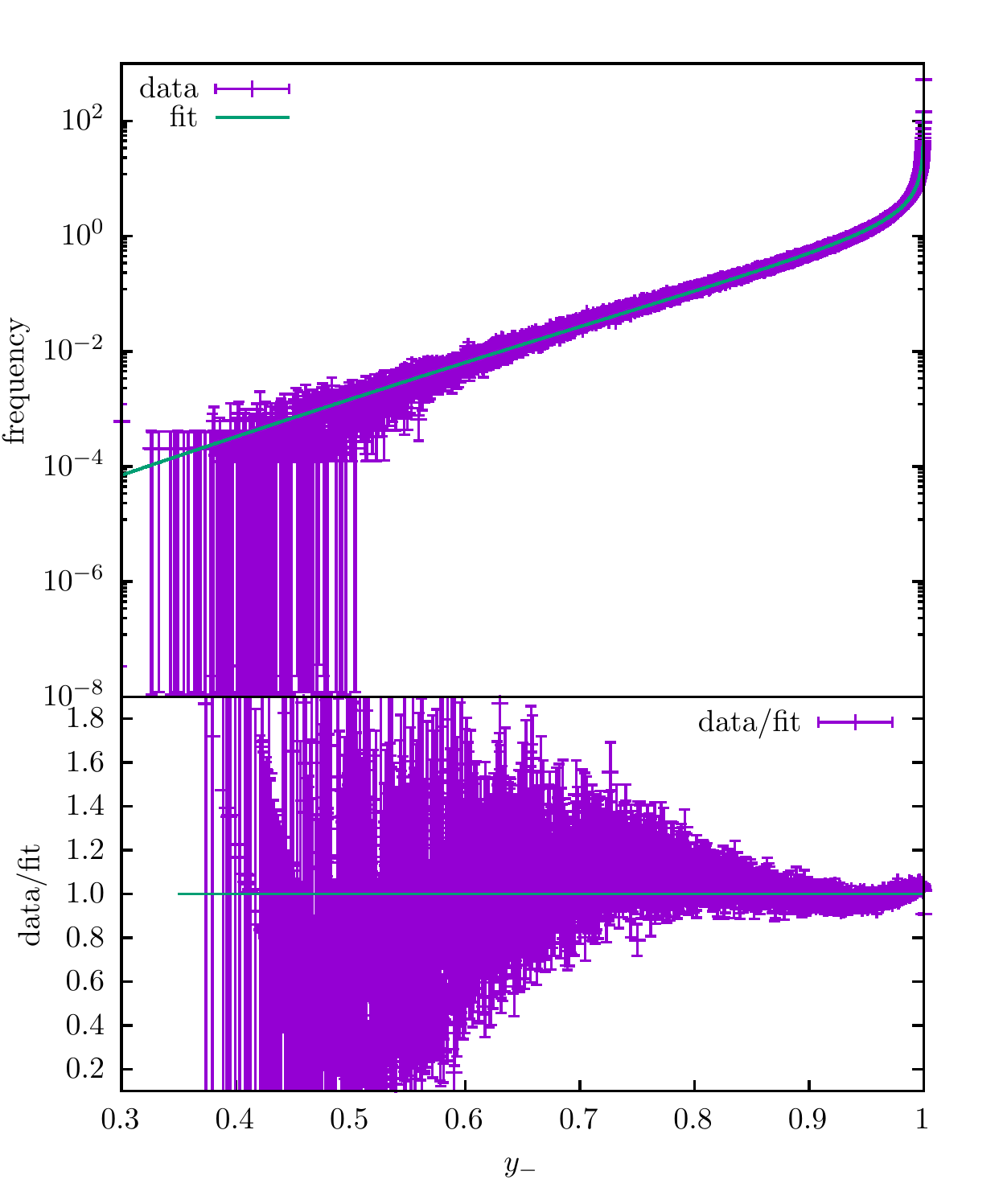}
\caption{\label{fig:f2dilc500}
As in fig.~\ref{fig:f2dfccee240}, for ILC500.
}
\end{center}
\end{figure}

\begin{figure}[th!]
\begin{center}
\includegraphics[width=0.45\textwidth]{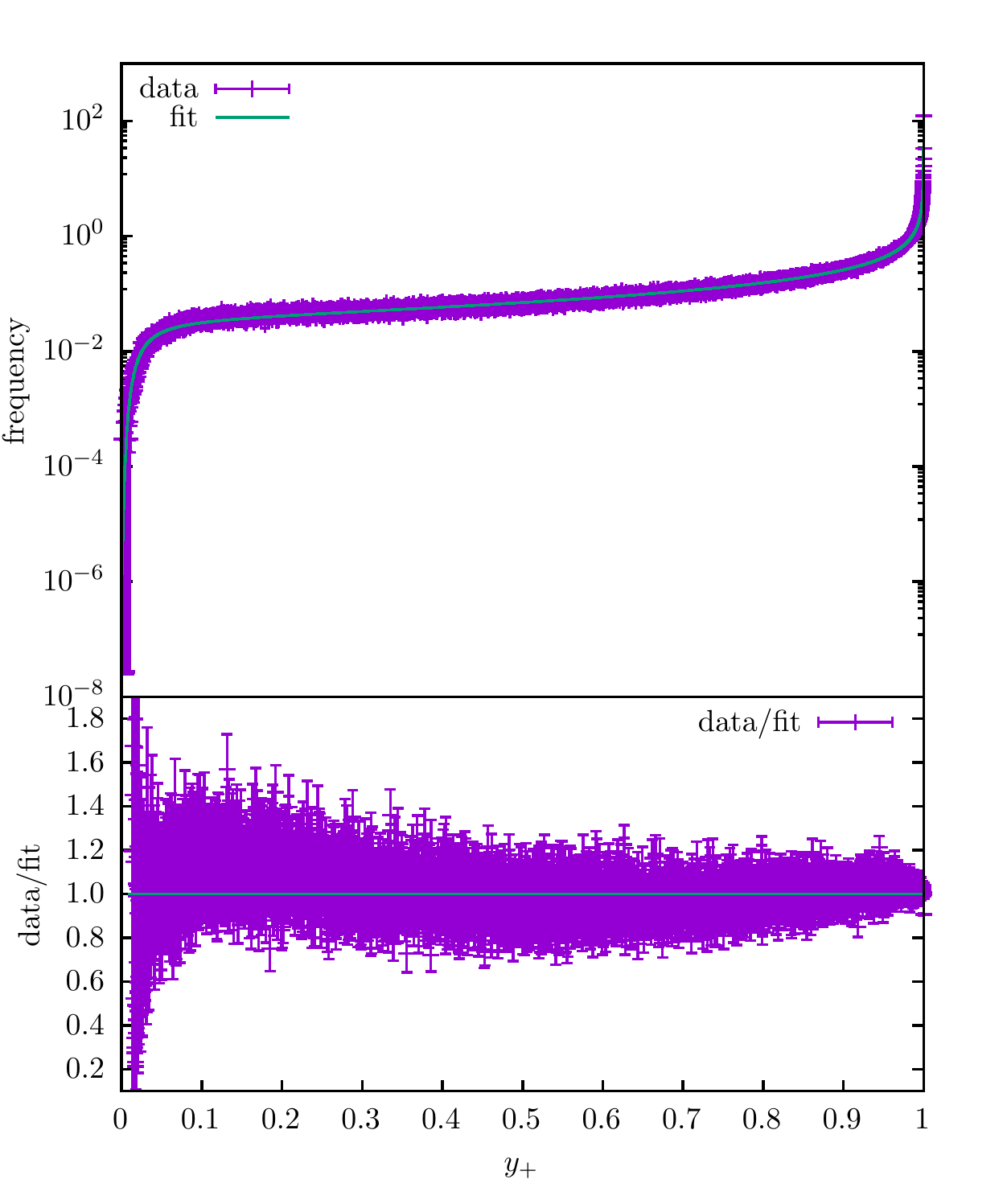}\hspace{5mm}
\includegraphics[width=0.45\textwidth]{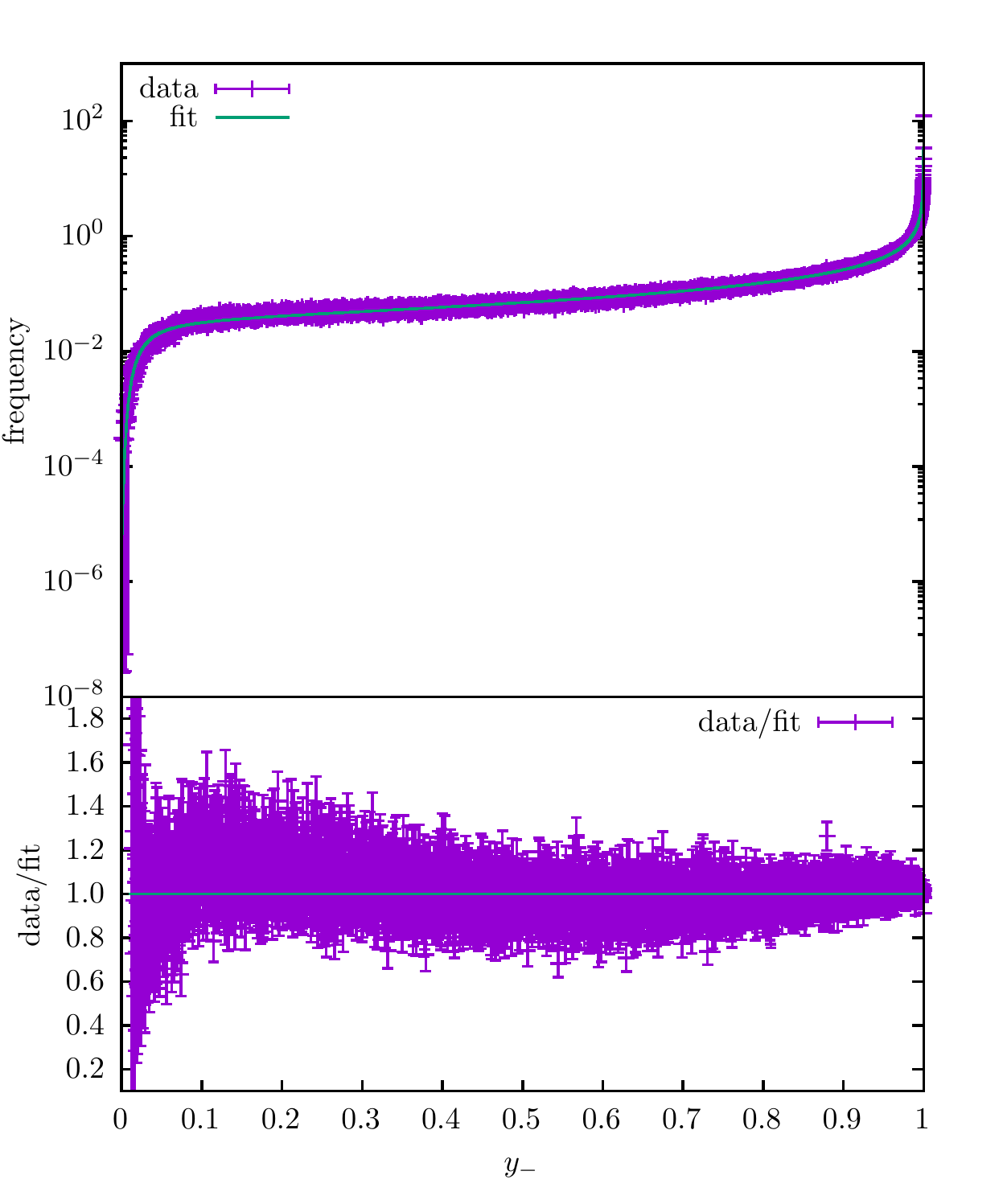}
\caption{\label{fig:f2dclic3000}
As in fig.~\ref{fig:f2dfccee240}, for CLIC3000.
}
\end{center}
\end{figure}

Turning to the cases of the linear-collider configurations
of table~\ref{tab:collpars}, the values of the parameters resulting
from the fits are reported in tables~\ref{tab:fit-ilc250}, 
\ref{tab:fit-ilc500}, and~\ref{tab:fit-clic3000} for ILC with 
$E_{beam}=125$~GeV (denoted by ILC250), ILC with $E_{beam}=250$~GeV 
(denoted by ILC500), and CLIC with $E_{beam}=1500$~GeV
(denoted by CLIC3000), respectively. The corresponding comparisons 
between the fitted functional forms and the simulation data are 
presented in appendix~\ref{sec:fitres}, in figs.~\ref{fig:f1dilc250}
and~\ref{fig:f2dilc250}, \ref{fig:f1dilc500} and~\ref{fig:f2dilc500}, 
and~\ref{fig:f1dclic3000} and~\ref{fig:f2dclic3000}.
\begin{table}[!htb]
\begin{center}
    \begin{tabular}{|c|c|c|c|c|c|c|}
        \hline
        $i$ & $\hat{f}_i$ & $p_i$ & $q_i$ & Integral\\
        \hline
        $11$ & 0.4933 &  &  &  \\
        \hline
        $01$ & 0.2085 & $-21.17$ & $-0.01452$ & 0.2016 \\
        \hline
        $10$ & 0.2085 & $-21.19$ & $-0.01460$ & 0.2016 \\
        \hline
        $00+$ & 0.3259 & $-20.22$ & $-0.01416$ & 0.3200\\
        $00-$ & 0.3263 & $-20.25$ & $-0.01420$ & 0.3203\\
        \hline
        Sum &  &  &  & 0.9990 \\
        \hline
    \end{tabular}
\end{center}
\caption{\label{tab:fit-ilc250}Fit results for ILC250.}
\end{table}
\begin{table}[!htb]
\begin{center}
    \begin{tabular}{|c|c|c|c|c|}
        \hline
        $i$ & $\hat{f}_i$ & $p_i$ & $q_i$ & Integral\\
        \hline
        $11$ & 0.5012 &  &  & \\
        \hline
        $01$ & 0.1613 & $-8.514$ & $-5.808$ & 0.1983\\
        \hline
        $10$ & 0.1613 & $-8.505$ & $-5.823$ & 0.1983\\
        \hline
        $00+$ & 0.2528 & $-7.535$ & $-6.790$ & 0.3173\\
        $00-$ & 0.2524 & $-7.481$ & $-6.849$ & 0.3171 \\
        \hline
        Sum &  &  &  & 0.9985 \\
        \hline
    \end{tabular}
\end{center}
\caption{\label{tab:fit-ilc500}Fit results for ILC500.}
\end{table}
\begin{table}
\begin{center}
    \begin{tabular}{|c|c|c|c|c|}
        \hline
        $i$ & $\hat{f}_i$ & $p_i$ & $q_i$ & Integral\\
        \hline
        $11$ & 0.7434 &  &  & \\
        \hline
        $01$ & 0.04563 & $-0.6016$ & $-0.007179$ & 0.1162\\
        \hline
        $10$ & 0.04568 & $-0.6089$ & $-0.007135$ & 0.1162 \\
        \hline
        $00+$ & 0.05943 & $-0.5629$ & $-0.007464$ & 0.1524 \\
        $00-$ & 0.05951 & $-0.5743$ & $-0.007259$ & 0.1523 \\
        \hline
        Sum &  &  &  & 0.9990 \\
        \hline
    \end{tabular}
\end{center}
\caption{\label{tab:fit-clic3000}Fit results for CLIC3000.}
\end{table}

We point out that the general functional form that we employ for
the fits at linear colliders is still that of eq.~(\ref{Bee2}), with
eq.~(\ref{kvals}) and eqs.~(\ref{f01fun})--(\ref{f00mfun}). However, 
while for ILC250 we adopt eq.~(\ref{ffuncirc}), for ILC500 and
CLIC3000 we use:
\beqn
f(y;p,q)&=&e^{p(1-y)}\,e^{q(1-y)^{3/2}}\,,
\label{ffunilc500}
\\
f(y;p,q)&=&e^{p(1-y)}\,e^{qy^{-3/2}}\,,
\label{ffunclic3000}
\eeqn
respectively, in view of the stronger beamstrahlung effects at these
colliders.

Owing to the symmetry between the two beams enforced by eq.~(\ref{Bee})
in the current implementation of \aNLOs, we could have decided to use
the same parameters for the functions relevant to the $\yp$ and $\ym$
variables, namely to set $\hat{f}_i=\hat{f}_j$, $p_i=p_j$, and
$q_i=q_j$ for $(i,j)=(01,10)$ and $(i,j)=(00+,00-)$ prior to fitting.
We have chosen not to do so, and rather verify that such identities
are (approximately) fulfilled post-fit; this constitutes a check on
the general correctness of the fitting procedure. Another check is
the unitarity of the luminosity, namely that the integral of the
${\cal B}_{\epem}(\yp,\ym)$ function must be equal to one\footnote{We
normalise the GuineaPig simulations to the total number of events.}.
In tables~\ref{tab:fit-fccee240}--\ref{tab:fit-clic3000} we report,
in the columns labelled as ``Integral'', the integrals of the
\mbox{$(1-\ypm)^{\kappa_\pm}f_i(\ypm)$} functions. The sum of the
result for $i=01$, plus that for $10$, plus the product of those for
$00+$ and $00-$, plus the value of $\hat{f}_{11}$, reported in the
tables as ``Sum'', gives the sought normalised luminosity. As we see,
that number is always very close to one.

The check on the unitarity of the luminosity guarantees that the fits 
behave as expected in the regions $\ypm\simeq 1$, which give by far
the dominant contributions to the physical cross sections. The quality
of the fits can also be assessed in a fully local way, by comparing
them to the simulation data, as is done in the figures we report
in appendix~\ref{sec:fitres}. As we see there, at circular colliders
there is a tendency for the fits to undershoot the data as small $\ypm$
values are approached. While this should be a minor problem in the
computation of observables (owing to the extremely small value assumed
by the beamstrahlung function in those regions), we stress that our
framework for fitting ${\cal B}_{\epem}(\yp,\ym)$ is fully flexible,
and can easily accomodate functions more involved than those of
eqs.~(\ref{ffuncirc}), (\ref{ffunilc500}), and~(\ref{ffunclic3000}).

\section{ISR and beamstrahlung\label{sec:ISRbeam}}
The separation of variables introduced in eq.~(\ref{Bexp}) allows one to 
define functions that include the parametrisation of both beamstrahlung and 
ISR effects. Indeed, it is straightforward to see that eqs.~(\ref{beamstr}) 
and~(\ref{factth0}) lead to\footnote{Equation~(\ref{factth1}) is fully 
general. In keeping with the characteristics of the present implementation 
(see eq.~(\ref{Bee2})), in the code we set \mbox{$(k,l)=(e^+,e^-)$}.}:
\beqn
d\Sigma_{\epem}(P_{\lp},P_{\lm})&=&\sum_{n=1}^N\sum_{ijkl}\int d\xp d\xm\,
\beamPDF{i}{k,n,kl}{\lp}(\xp,\mu^2,m^2)\,
\beamPDF{j}{l,n,kl}{\lm}(\xm,\mu^2,m^2)
\nonumber\\*&&\phantom{\sum_{ij}\int}\times
d\hsig_{ij}(\xp P_{\lp},\xm P_{\lm},\mu^2,m^2)\,,
\label{factth1}
\eeqn
where:
\beq
\beamPDF{i}{k,n,kl}{\lpm}(x,\mu^2,m^2)=
\int dy\,dz\,\delta(x-yz)\,b^{(\lpm)}_{n,kl}(y)\,
\PDF{i}{k}(z,\mu^2,m^2)\,.
\label{beamPDF}
\eeq
While eq.~(\ref{factth1}) is equivalent to the combination of
eqs.~(\ref{beamstr}) and~(\ref{factth0}), it gives one the possibility
of computing a collider-level cross section in an alternative way
w.r.t.~to the latter two. This is because the quantities introduced
in eq.~(\ref{beamPDF}) depend only on the collider type
(through the beamstrahlung function) and on the small-angle initial
state radiation (through the PDFs). In other words, they are process
{\em in}dependent, and as such can be parametrised once and for all.

We exploit this fact in the following way. Firstly, the leading $x\to 1$
behaviour of the functions on the l.h.s.~of eq.~(\ref{beamPDF}) can be 
analytically computed if the same kind of information is available for 
the functions in the integrand on the r.h.s.~of eq.~(\ref{beamPDF}),
which is the case (see eq.~(\ref{Bee2}) and 
refs.~\cite{Bertone:2019hks,Frixione:2021wzh}).
The worst-case scenario is again that relevant to electron/positron
PDFs, as was already pointed out in sect.~\ref{sec:ISR}. By using
eq.~(\ref{eePDFasy}) one immediately arrives at:
\beqn
\beamPDF{\lpm}{\lpm,n,\epem}{\lpm}(x)&\stackrel{x\to 1}{\longrightarrow}&
\frac{\hbeamPDF{\lpm}{k,n,\epem}{\lpm}(x)}{(1-x)^{1-\beta}}
\;\;\;\;\;\;\;\;{\rm if}\;\;\;\;\;\;\;\;
b^{(\lpm)}_{n,\epem}(y)\propto\delta(1-y)\,,
\label{phiasy1}
\\*
\beamPDF{\lpm}{\lpm,n,\epem}{\lpm}(x)&\stackrel{x\to 1}{\longrightarrow}&
\frac{\hbeamPDF{\lpm}{k,n,\epem}{\lpm}(x)}{(1-x)^{-\kappa-\beta}}
\;\;\;\;\;\;\;\;{\rm if}\;\;\;\;\;\;\;\;
b^{(\lpm)}_{n,\epem}(y)\propto (1-y)^\kappa\,.
\label{phiasy2}
\eeqn
Thus, we introduce the analogue of eq.~(\ref{eePDFint}) as follows:
\beq
\beamPDF{i}{k,n,kl}{\lpm}(x)=
\frac{\obeamPDF{i}{k,n,kl}{\lpm}(x)}{(1-x)^{\bkappa}}\,,
\label{obeampdf}
\eeq
and construct routines that, given the beamstrahlung functions
and the PDFs, return the $\overline{\phi}_\alpha$ functions and
the coefficient $\bkappa$. In order to do this, we need to bear in mind
the discussion presented in sect.~\ref{sec:ISR} about the $z\to 1$
behaviour of the PDFs, and the necessity of employing the free
parameter $\gamma$, in order to be able put the LO+LL and NLO+NLL
PDFs on the same footing. For the examples of eqs.~(\ref{phiasy1})
and~(\ref{phiasy2}), we have:
\beqn
{\rm eq}.~(\protect\ref{phiasy1})
\phantom{aaa}&\longrightarrow&\phantom{aaa}
\bkappa=\gamma\,,
\\*
{\rm eq}.~(\protect\ref{phiasy2})
\phantom{aaa}&\longrightarrow&\phantom{aaa}
\bkappa=\gamma-1-\kappa\,.
\eeqn
For what concerns the functions $\overline{\phi}_\alpha$, the integral
in eq.~(\ref{beamPDF}) is computed for several pre-defined values 
of the pair \mbox{$(x,\mu^2)$} -- such values thus constitute
the nodal points of a two-dimensional grid. The integral results are
written in a file, together with a routine that reads them and, by using
the information on the nodal points as well, returns the sought functions
for any arbitrary values of $x$ and $\mu^2$ by performing a bi-linear 
interpolation of the stored results. Since the functions 
$\overline{\phi}_\alpha$ are well-behaved in the whole $x$ range
(thanks to the factor \mbox{$(1-x)^{\bkappa}$} introduced in 
eq.~(\ref{obeampdf})), the procedure just outlined is sufficient
to achieve a satisfactory level of precision. In particular, for the 
numerical tests performed so far, which are restricted to LO+LL results,
we have used the nodal points obtained with 100 $x$ values (distributed 
according to a polynomial law that accumulates relatively more points 
towards $x=0$ and $x=1$) and 20 $\mu^2$ values (distributed linearly 
in $\log\mu^2$ in the range \mbox{$\mu\in [1,10^4]$~GeV}), but we stress 
that these are input parameters to
the code that constructs the interpolating grids, and can thus be 
changed at will. Such a code will be distributed with \aNLOs, in
order to allow the user to constructs his/her own $\overline{\phi}_\alpha$
functions specific for the $\epem$ collider of interest.

Having parametrised once and for all the functions $\beamPDF{i}{k,n,kl}{\lp}$,
eq.~(\ref{factth1}) is exactly on the same footing as eq.~(\ref{factth0}),
and therefore the same strategy described in sect.~\ref{sec:ISR} for
the integration of the particle cross section can be used to obtain
the collider-level cross section. In principle, this observation 
applies to each of the summands in the index $n$ that appear in
eq.~(\ref{factth1}). However, we have so far found no loss of accuracy
by treating all of those terms with the same change of variables, which
is dictated by the worst-case scenario $\bkappa=\gamma$. The advantage of
this is that one can use the luminosity function
\beq
\sum_{n=1}^N\,
\beamPDF{i}{k,n,kl}{\lp}(\xp,\mu^2,m^2)\,
\beamPDF{j}{l,n,kl}{\lm}(\xm,\mu^2,m^2)
\label{beamlumi}
\eeq
as a single number that multiplies that partonic cross section
(in view of the fact that the latter is $n$-independent); this implies
that the number of integration channels for a collider-level cross section
is the same as that relevant to the corresponding particle cross section.
Having said that, we stress that the separate treatment of the contributions
to eq.~(\ref{factth1}) associated with different $n$ values is a pure matter 
of implementation, without any conceptual implications. 

We conclude this section with an observation on the calculation of
collider-level cross sections. Equation~(\ref{factth1}) is identical
to eq.~(\ref{beamstr}), where the particle cross section in the latter
equation is computed with eq.~(\ref{factth0}). However, from the numerical
viewpoint eq.~(\ref{beamstr}) entails the generation of two extra
integration variables w.r.t.~what happens with eq.~(\ref{factth1}) --
with the former, one needs to generate $\ypm$ and $\zpm$, while for
the latter only $\xpm$ are generated. This happens thanks to the 
pre-tabulation of the $\phi_\alpha$ functions; in other words, one
trades the generation of two integration variables in the context
of cross-sections runs for the generation of two variables in the
pre-tabulation phase, i.e.~prior to physics runs. This is clearly 
helpful in reducing the complexity of the calculations, and thus
in achieving a faster convergence for a given accuracy target.
The only possible drawback of adopting eq.~(\ref{factth1}) is that 
by doing so the event-by-event information on the incoming-particle
energies are lost; they can however be recovered on statistical basis,
by randomly generating (for each of two incoming legs) the variable $y$ 
that appear in eq.~(\ref{beamPDF}) according to its distribution as defined 
by the integrand on the r.h.s.~of that equation.

\section{Running the code\label{sec:run}}
The generic instructions to run \aNLOs\ have been described elsewhere 
(see e.g.~refs.~\cite{Alwall:2014hca,Frederix:2018nkq}). Here we limit
ourselves to discussing those specific to lepton-lepton collisions. 

The process generation and output stages are unchanged w.r.t.~the current 
usage. The code can be run in either the LO or the NLO-QCD mode; the NLO-EW
mode is not yet supported. At the NLO in QCD, both fixed-order (fNLO) and
parton-shower matched computations (NLO+PS) are supported; in the latter
case, the user may also enable QED showers (this does not imply any QED
matching, given that the underlying matrix elements are restricted to 
being of LO in the EW coupling constant). Prior to running the code,
the user must specify in {\tt run\_card.dat} that ISR and possibly 
beamstrahlung are to be included. This is done by means of the {\tt lpp1}
and {\tt lpp2} variables, which must be set equal to $+3$ ($-3$) for an
electron (positron) beam. Parameterisations of ISR (plus possibly 
beamstrahlung) effects, as described in 
sects.~\ref{sec:ISR}--\ref{sec:ISRbeam}, 
are provided as tabulated fortran files, stored in the directories 
{\tt Source/PDF/lep\_densities/XXX/}. Here {\tt XXX} denotes a specific 
choice of ISR(+beamstrahlung) parametrisation. The options included in
the code release are {\tt XXX=isronlyll} (for the LL+LO electron/positron 
PDFs), {\tt cepc240ll}, {\tt clic3000ll}, {\tt fcce240ll}, {\tt fcce365ll}, 
and {\tt ilc500ll} (for the beamstrahlung configurations described in 
sect.~\ref{sec:beamstr} convoluted with the LL+LO PDFs). For reasons of
benchmarking and compatibility with results in the literature, the
following form for the LL+LO electron/positron PDFs is presently adopted:
\beqn
\PDF{\lpm}{\lpm}(z)&=&
\frac{e^{3\beta/4-\gamma_{\sss\rm E}\beta}}{\Gamma\left(1+\beta\right)}
\beta(1-z)^{\beta-1}-\frac{\beta}{2}\,h_1(z)-\frac{\beta^2}{8}\,h_2(z)\,,
\\
h_1(z)&=&1+z\,,
\\
h_2(z)&=&\frac{1+3z^2}{1-z}\ln(z)+4(1+z)\ln(1-z)+5+z\,,
\eeqn
where
\beq
\beta=\frac{\aem}{\pi}\left(\log\frac{\mu^2}{m^2}-1\right)\,.
\eeq
At the later stage, we shall implement all options stemming from
refs.~\cite{Bertone:2019hks,Frixione:2021wzh}.
The user's choice of the ISR(+beamstrahlung) parametrisation are set
in {\tt run\_card} by means of the assignment {\tt pdlabel=XXX}. 
New ISR(+beamstrahlung) scenarios can be added by creating a new directory
under {\tt Source/PDF/lep\_densities} as discussed before; this must contain
exactly two files named {\tt eepdf.f} and {\tt gridpdfaux.f}, using the 
formats of those present in one of the directories already available.

\section*{Acknowledgments}
M.Z.~is supported by the ``Programma per Giovani Ricercatori Rita Levi
Montalcini'', granted by the Italian Ministero dell'Universit\`a e 
della Ricerca (MUR). X.Z.~is supported by the Italian Ministry of Research 
(MUR) under grant PRIN 20172LNEEZ. The work of O.M.~and of X.~Z.~has been 
supported in part by the European Union's Horizon 2020 research and innovation 
programme as part of the Marie S\l{}odowska-Curie Innovative Training Network 
MCnetITN3 (grant agreement no. 722104). The authors thank Fabio Maltoni
for comments on the manuscript.

\appendix
\section{Fit results\label{sec:fitres}}
In this appendix we complement the information, reported in
sect.~\ref{sec:beamstr}, on the fits to the beamstrahlung function,
by comparing the results of such fits with the data whence they stem.

For each collider configuration considered in sect.~\ref{sec:beamstr}
there are two figures (that correspond to the second and third line of 
eq.~(\ref{Bee2}), and to the fourth line of that equation, respectively), 
each of which has two panels (that corrrespond to the $\yp$ and $\ym$ 
variables). All panels have the same layout, composed of a main frame 
and a lower inset (note, however, that the domain and codomain ranges 
are generally different across the various figures). In the main frame 
the GuineaPig data (including their error bars; data are normalised to 
their total number, hence the label ``frequency''), are displayed as 
magenta points, while the fit result is shown as a piecewise continuous 
green line; the inset shows the ratio of the quantities that appear in 
the main frame, in the form data over fit.

\begin{figure}[th!]
\begin{center}
\includegraphics[width=0.45\textwidth]{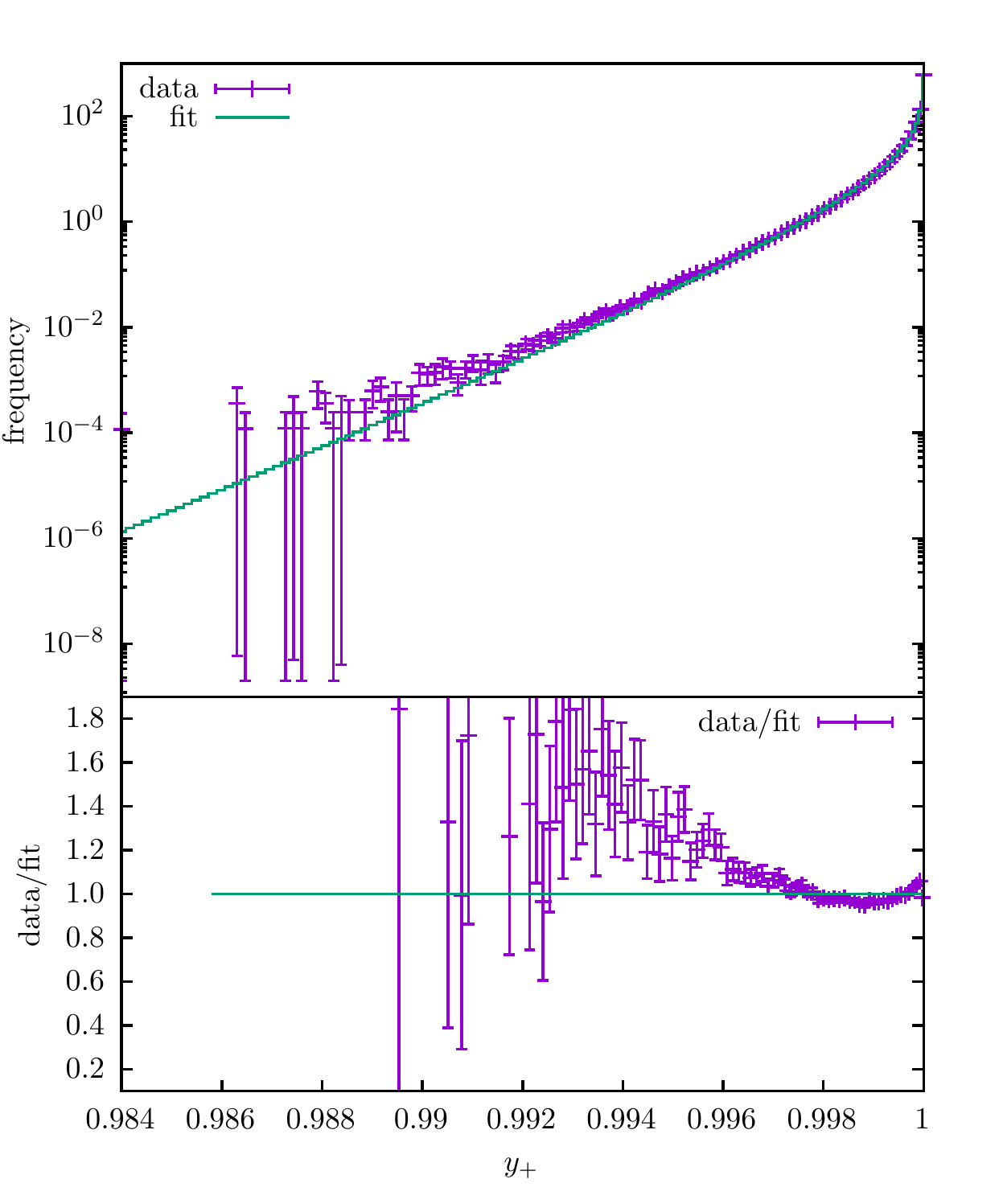}\hspace{5mm}
\includegraphics[width=0.45\textwidth]{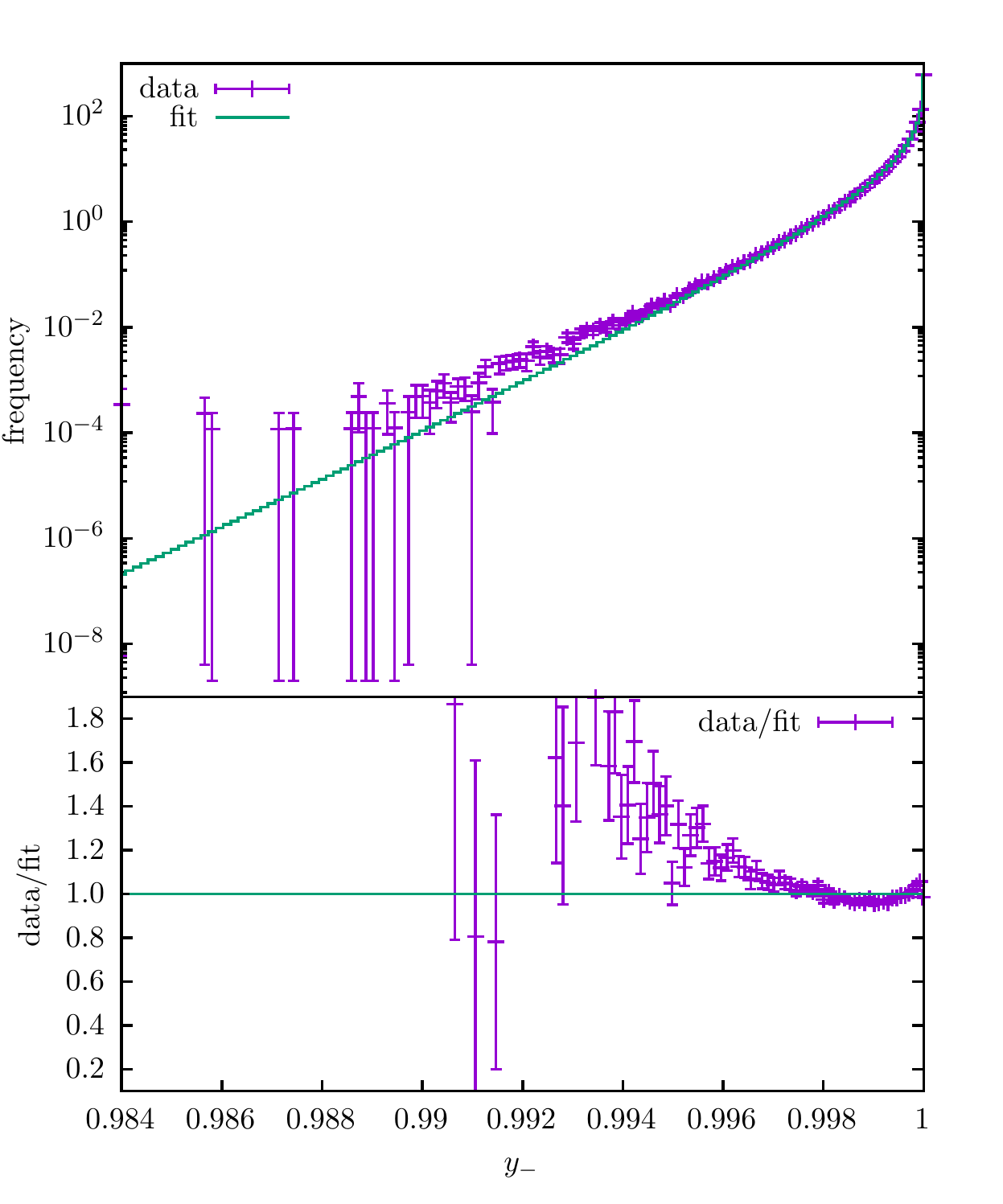}
\caption{\label{fig:f1dfccee240}
Fits to the second (left panel) and third (right panel) of eq.~(\ref{Bee2}),
for FCC-ee240.
}
\end{center}
\end{figure}

\begin{figure}[th!]
\begin{center}
\includegraphics[width=0.45\textwidth]{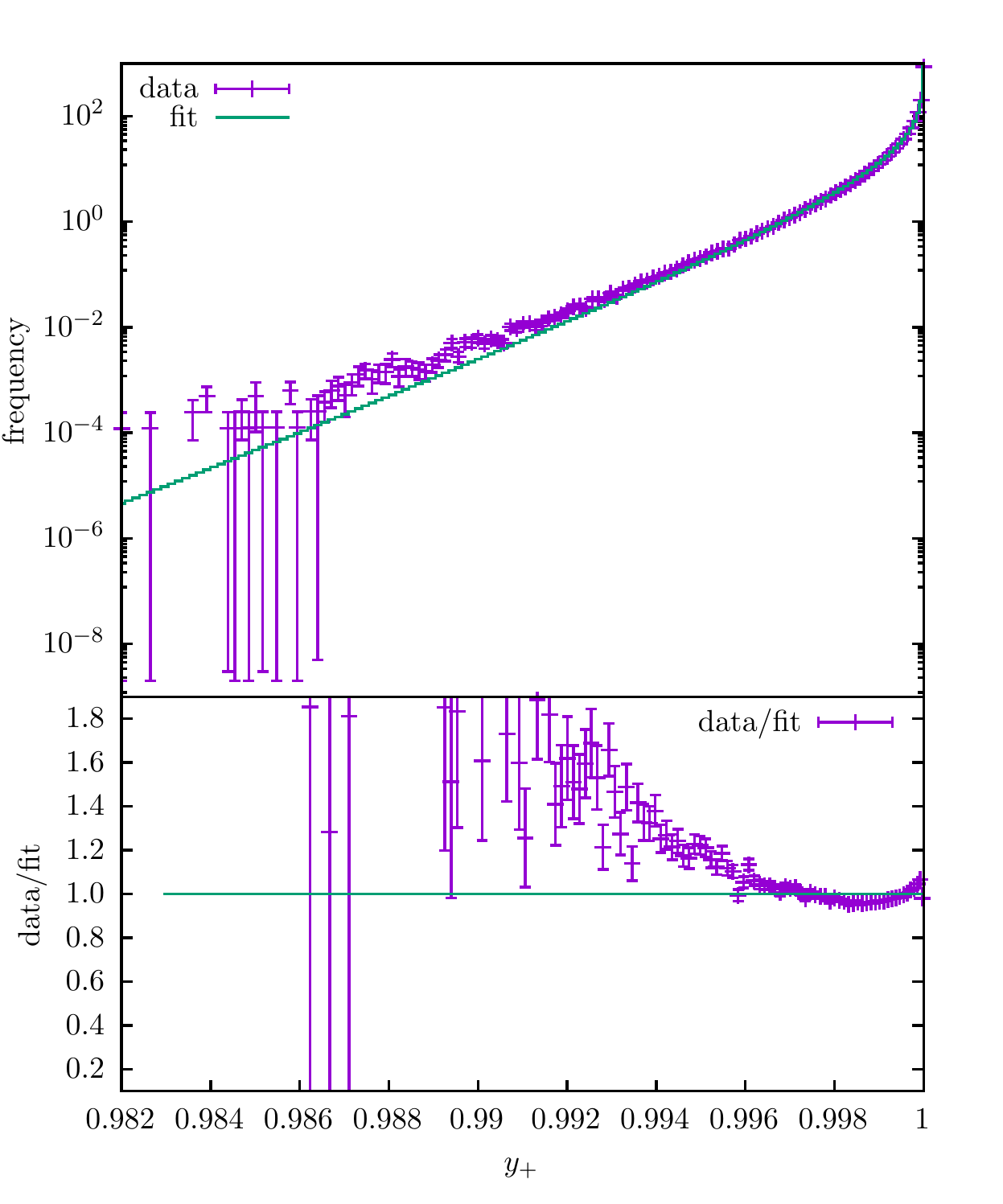}\hspace{5mm}
\includegraphics[width=0.45\textwidth]{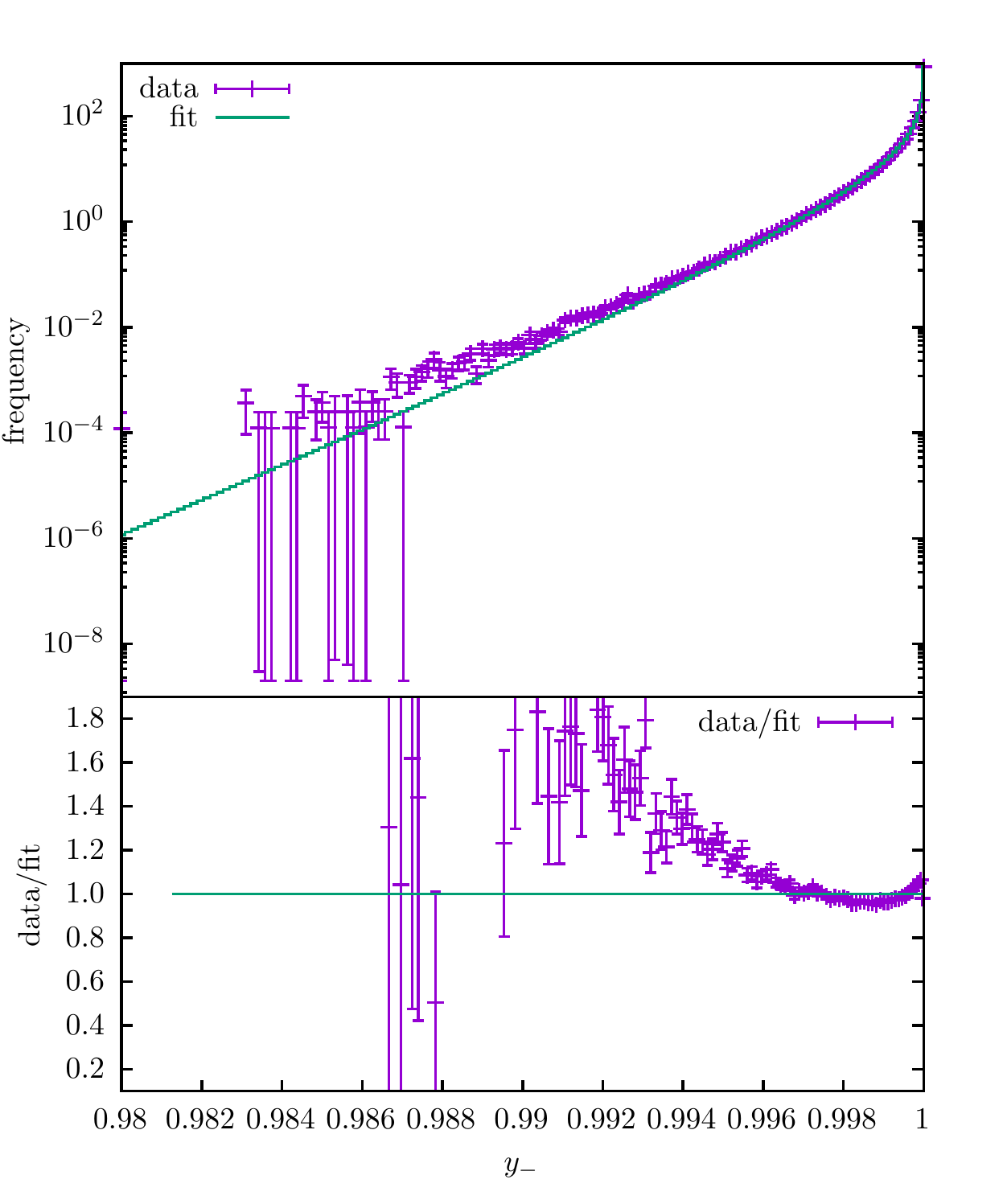}
\caption{\label{fig:f1dfccee365}
As in fig.~\ref{fig:f1dfccee240}, for FCC-ee365.
}
\end{center}
\end{figure}

\begin{figure}[th!]
\begin{center}
\includegraphics[width=0.45\textwidth]{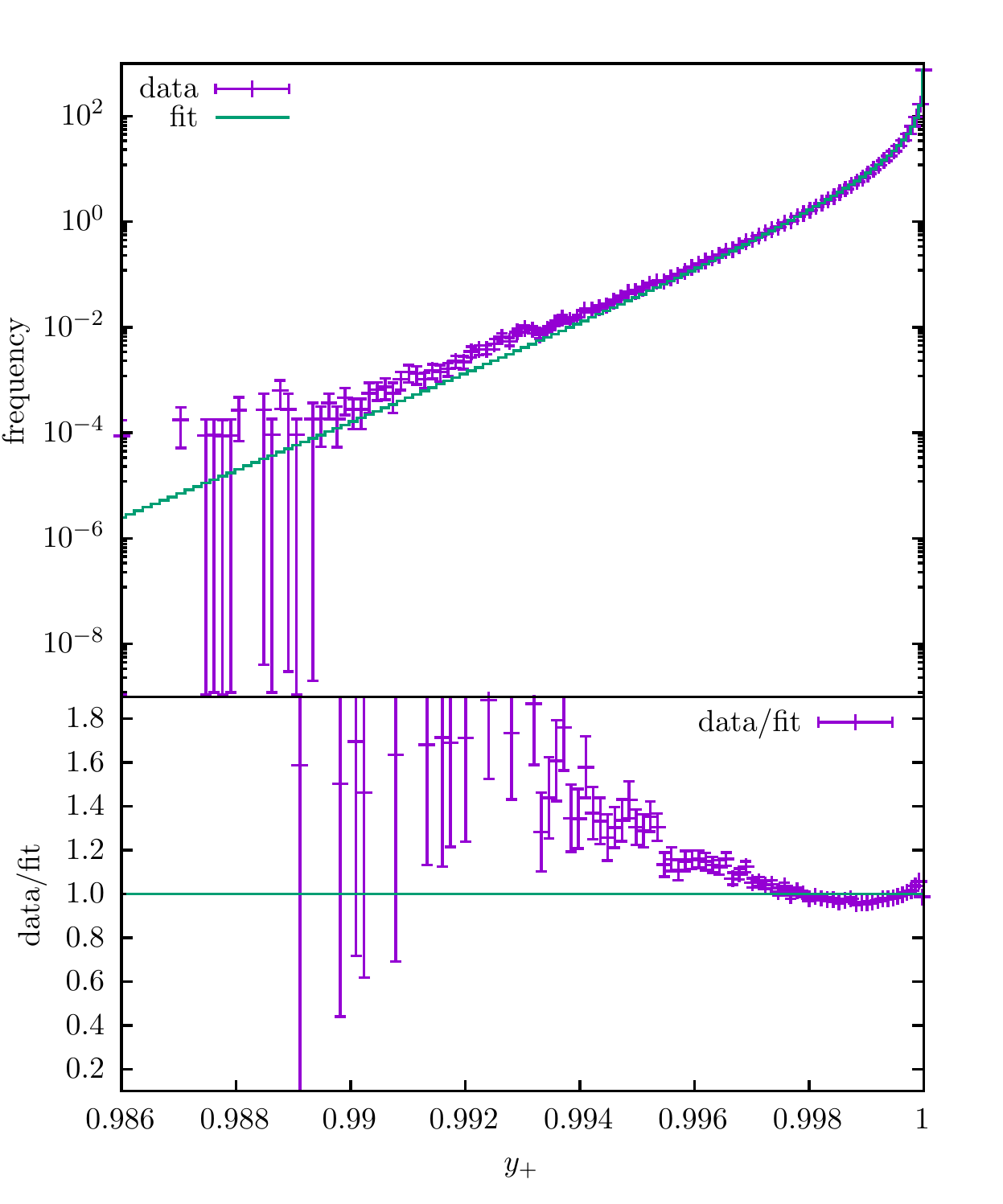}\hspace{5mm}
\includegraphics[width=0.45\textwidth]{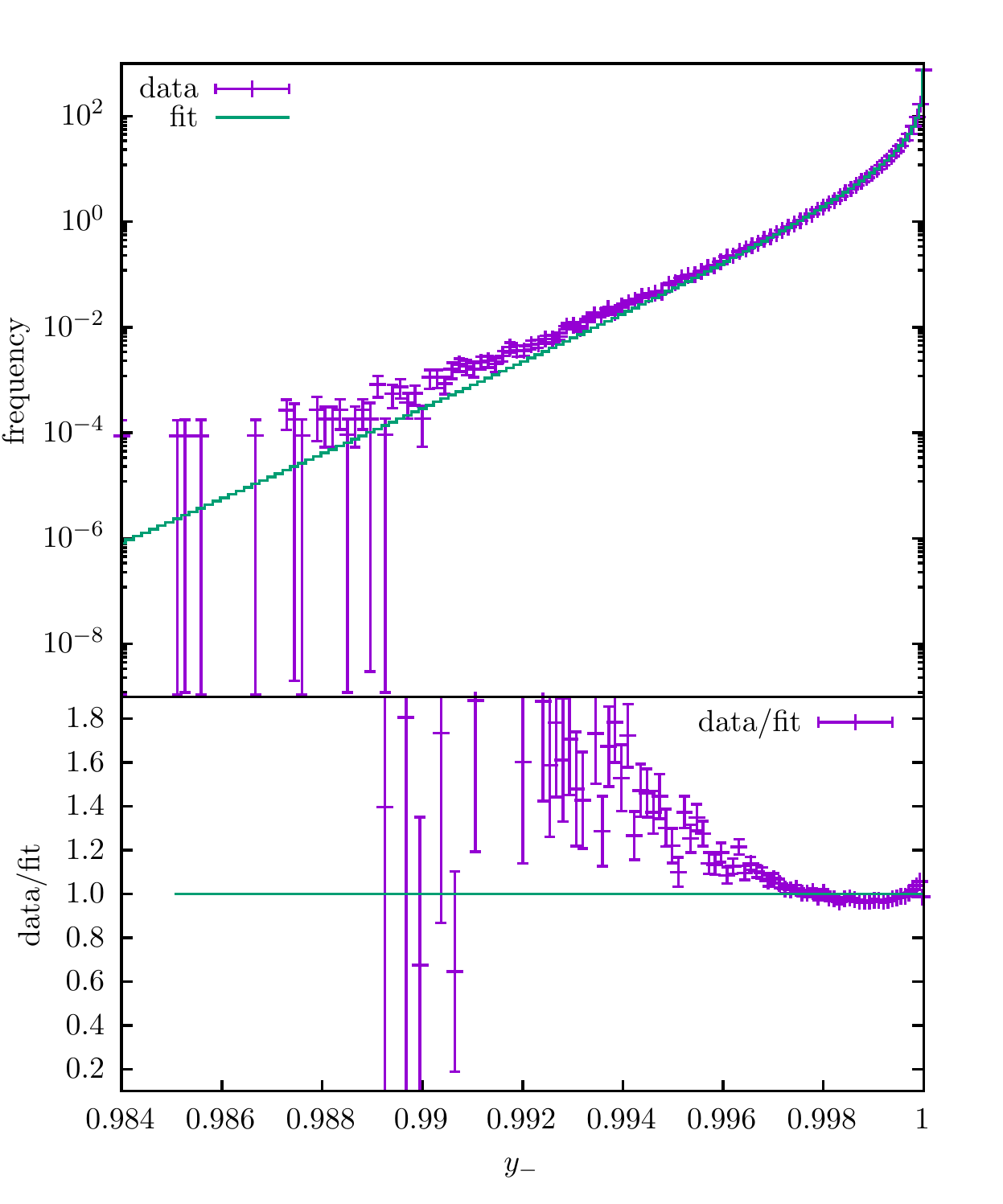}
\caption{\label{fig:f1dcepc240}
As in fig.~\ref{fig:f1dfccee240}, for CEPC240.
}
\end{center}
\end{figure}

\begin{figure}[th!]
\begin{center}
\includegraphics[width=0.45\textwidth]{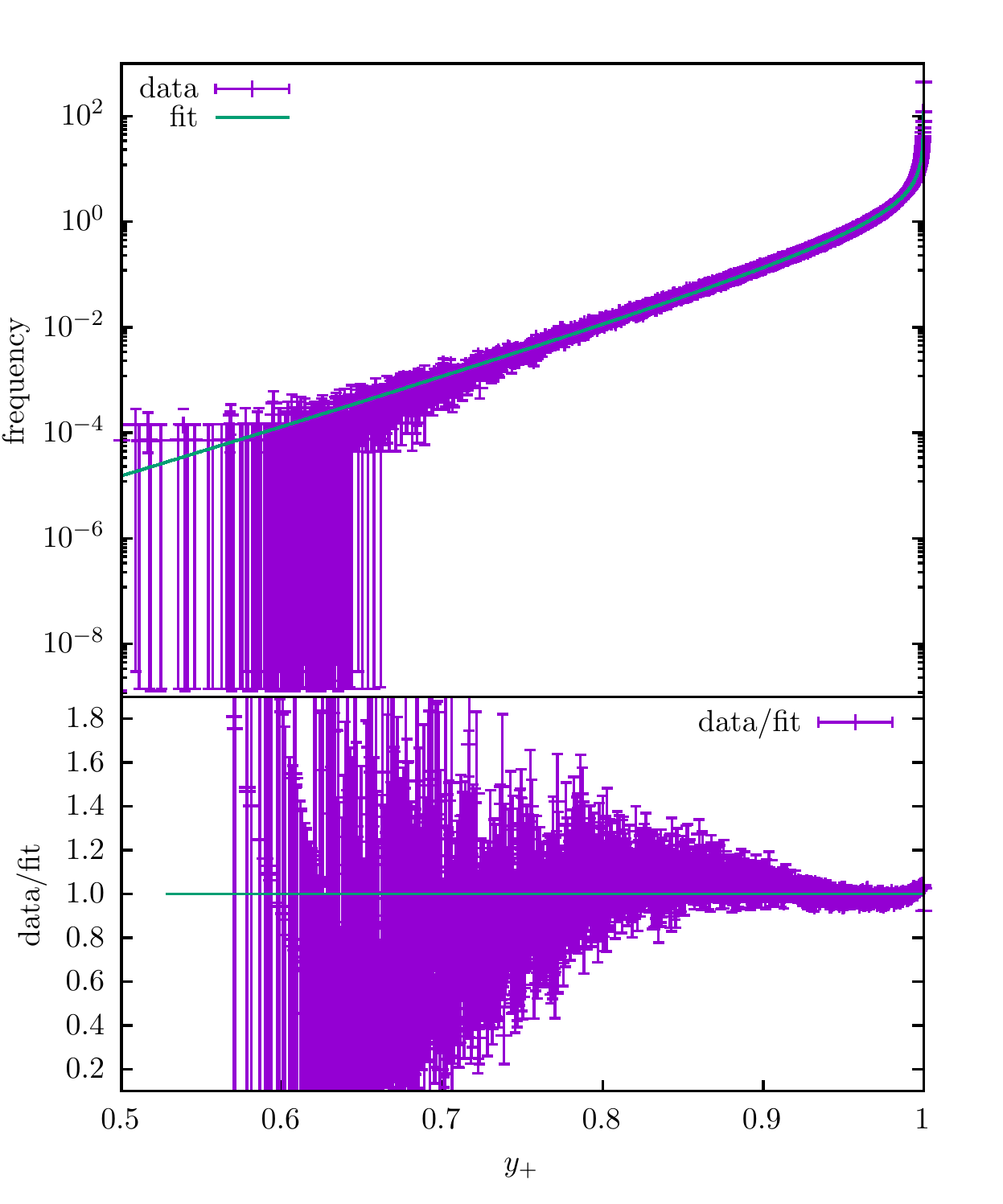}\hspace{5mm}
\includegraphics[width=0.45\textwidth]{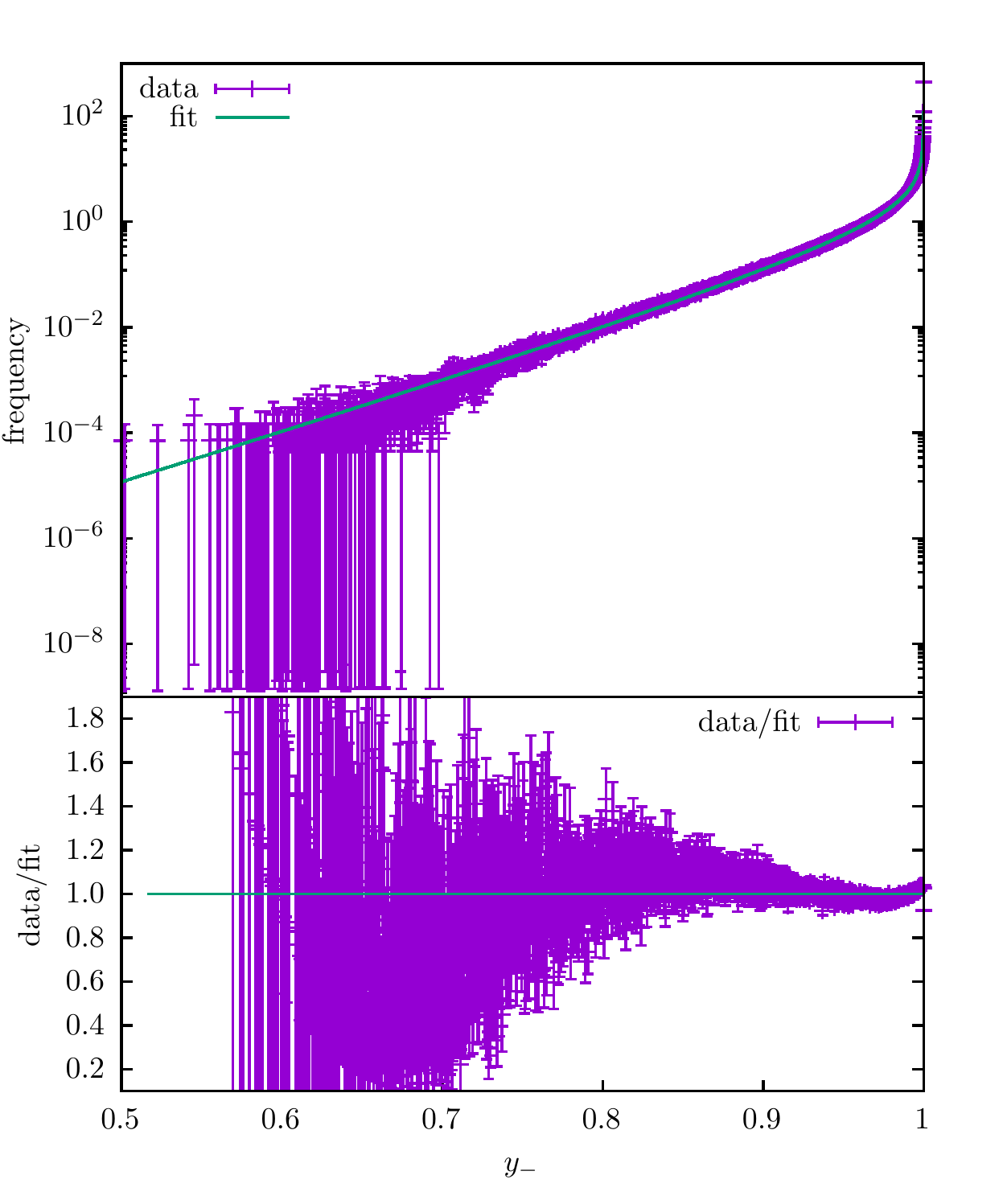}
\caption{\label{fig:f1dilc250}
As in fig.~\ref{fig:f1dfccee240}, for ILC250.
}
\end{center}
\end{figure}

\begin{figure}[th!]
\begin{center}
\includegraphics[width=0.45\textwidth]{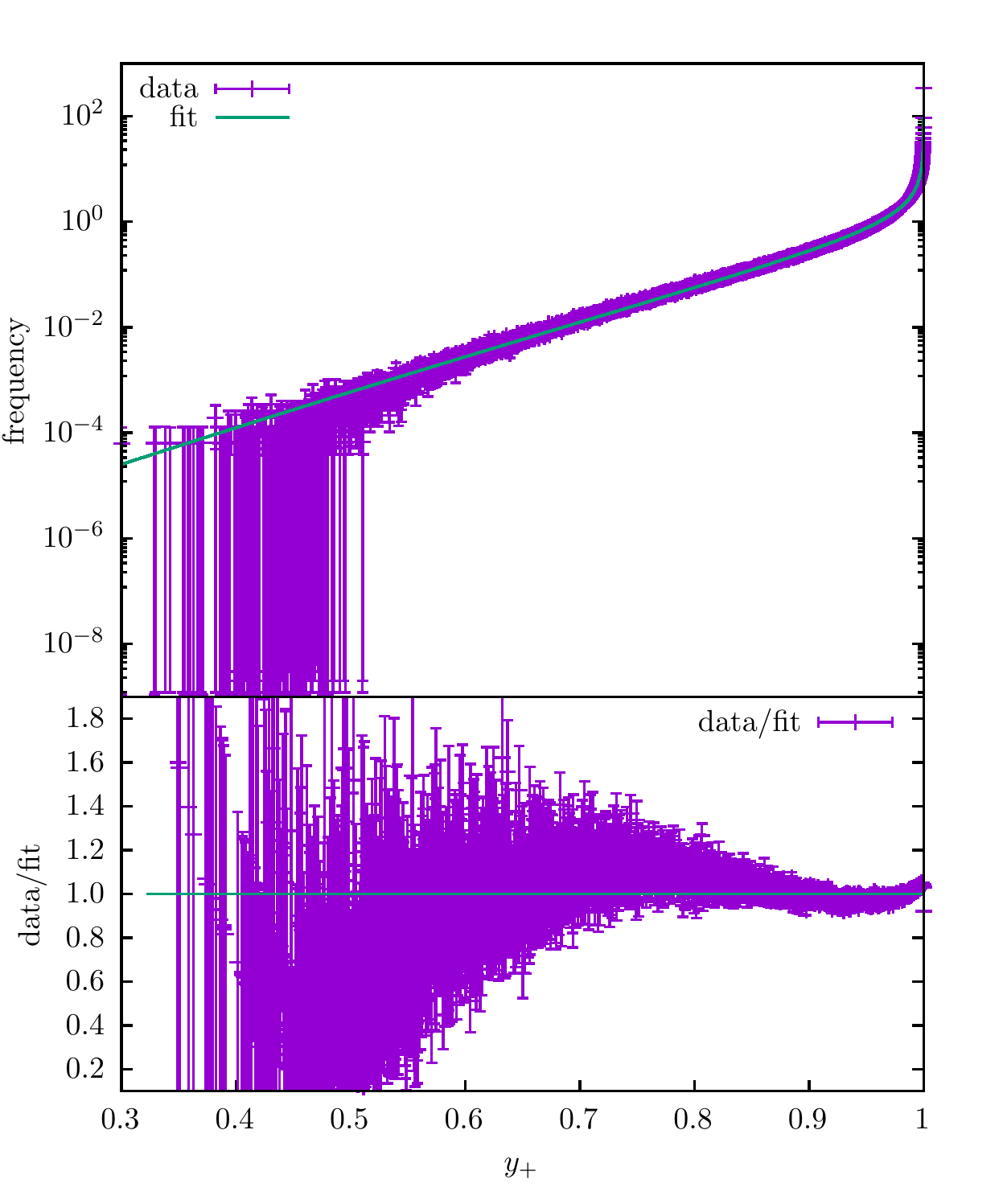}\hspace{5mm}
\includegraphics[width=0.45\textwidth]{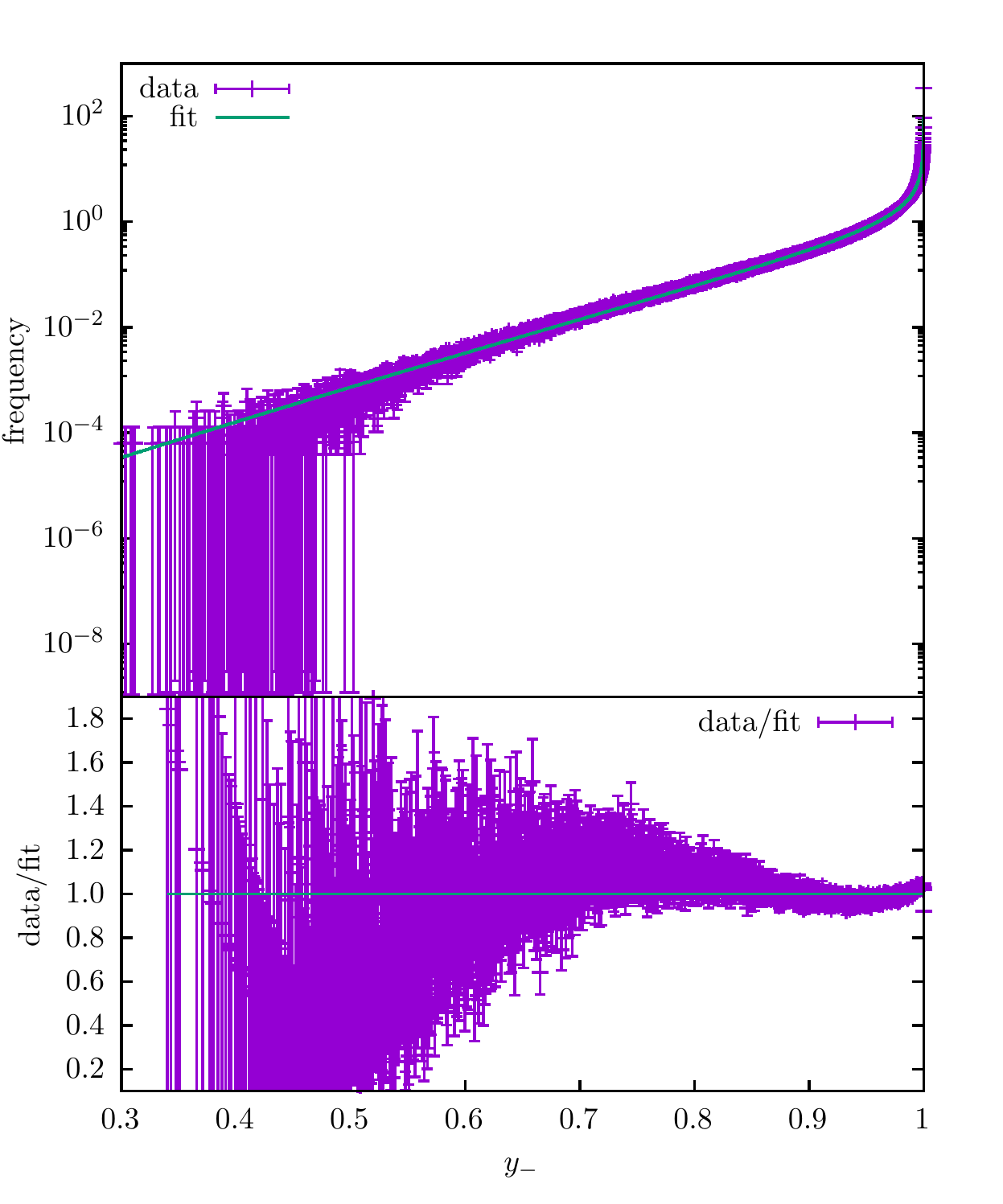}
\caption{\label{fig:f1dilc500}
As in fig.~\ref{fig:f1dfccee240}, for ILC500.
}
\end{center}
\end{figure}

\begin{figure}[th!]
\begin{center}
\includegraphics[width=0.45\textwidth]{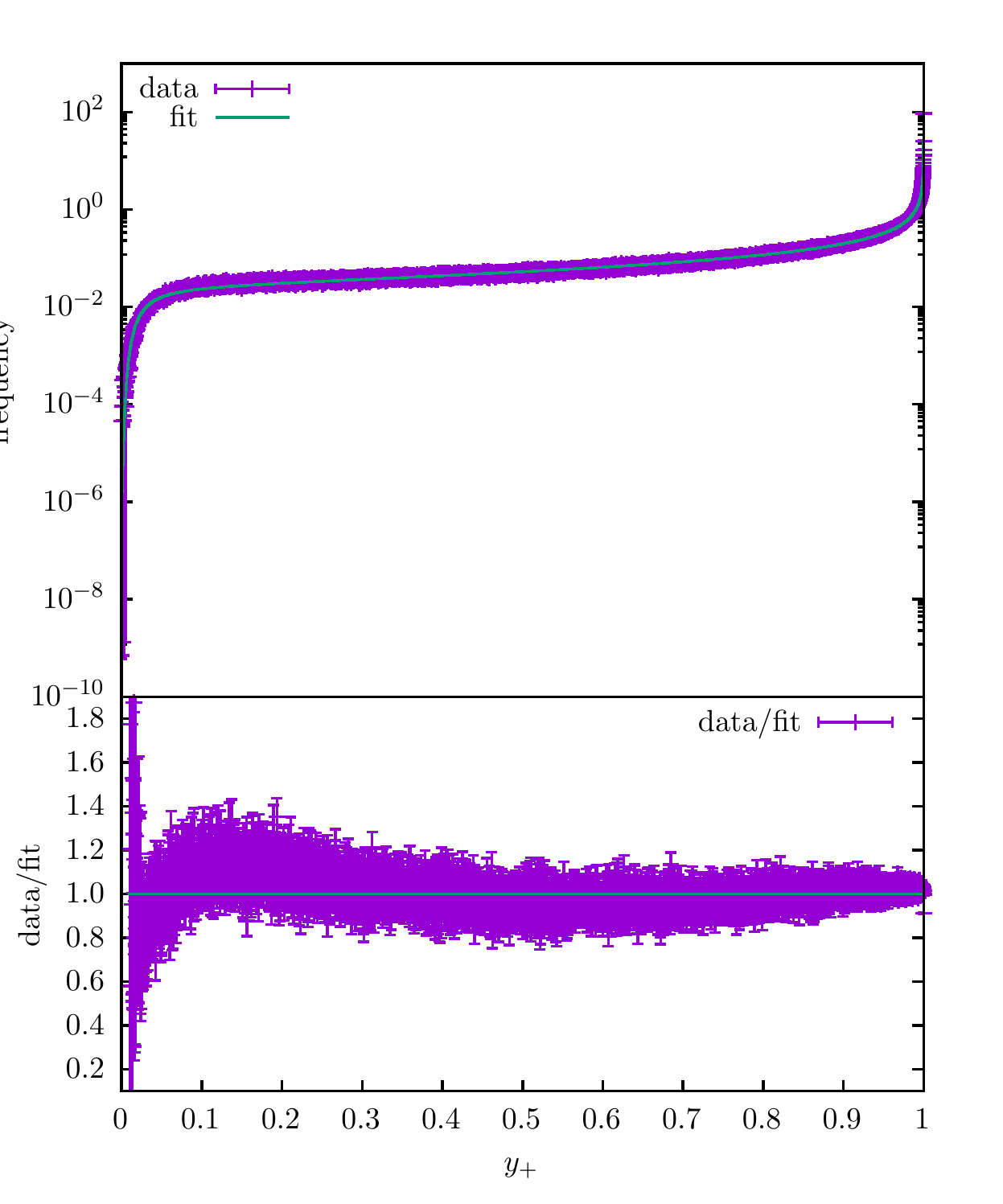}\hspace{5mm}
\includegraphics[width=0.45\textwidth]{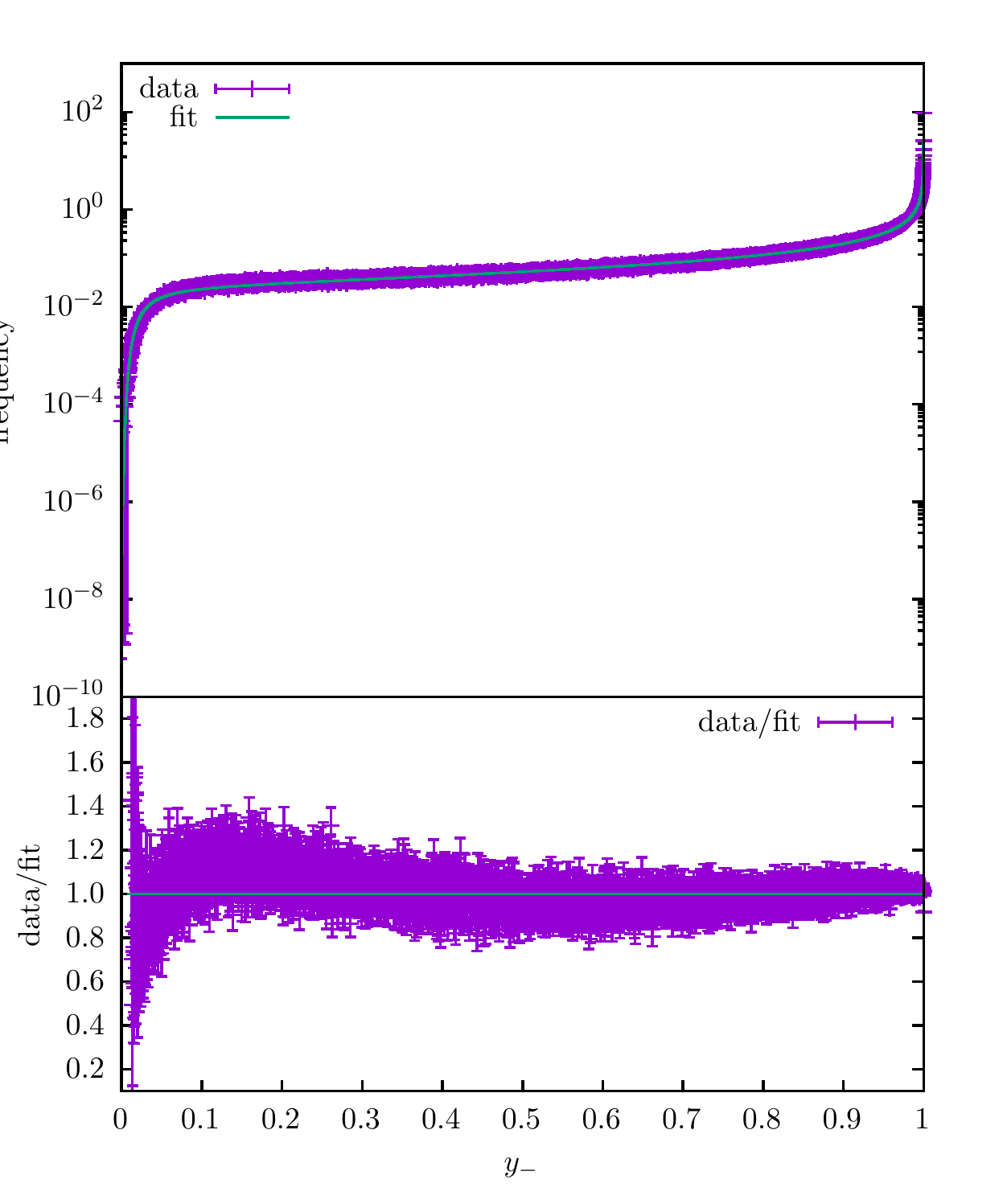}
\caption{\label{fig:f1dclic3000}
As in fig.~\ref{fig:f1dfccee240}, for CLIC3000.
}
\end{center}
\end{figure}

\bibliographystyle{JHEP}
\bibliography{eeLO}

\providecommand{\href}[2]{#2}\begingroup\raggedright\begin{thebibliography}{10}

\bibitem{Alwall:2014hca}
J.~Alwall, R.~Frederix, S.~Frixione, V.~Hirschi, F.~Maltoni, O.~Mattelaer
  et~al., \emph{{The automated computation of tree-level and next-to-leading
  order differential cross sections, and their matching to parton shower
  simulations}}, \href{http://dx.doi.org/10.1007/JHEP07(2014)079}{\emph{JHEP}
  {\bf 07} (2014) 079}, [\href{http://arxiv.org/abs/1405.0301}{{\tt
  1405.0301}}].

\bibitem{Degrande:2011ua}
C.~Degrande, C.~Duhr, B.~Fuks, D.~Grellscheid, O.~Mattelaer et~al., \emph{{UFO
  - The Universal FeynRules Output}},
  \href{http://dx.doi.org/10.1016/j.cpc.2012.01.022}{\emph{Comput.Phys.Commun.}
  {\bf 183} (2012) 1201--1214}, [\href{http://arxiv.org/abs/1108.2040}{{\tt
  1108.2040}}].

\bibitem{Alwall:2006yp}
J.~Alwall, A.~Ballestrero, P.~Bartalini, S.~Belov, E.~Boos et~al., \emph{{A
  Standard format for Les Houches event files}},
  \href{http://dx.doi.org/10.1016/j.cpc.2006.11.010}{\emph{Comput.Phys.Commun.}
  {\bf 176} (2007) 300--304}, [\href{http://arxiv.org/abs/hep-ph/0609017}{{\tt
  hep-ph/0609017}}].

\bibitem{Butterworth:2010ym}
J.~Butterworth, A.~Arbey, L.~Basso, S.~Belov, A.~Bharucha et~al., \emph{{The
  Tools and Monte Carlo working group Summary Report}},
  \href{http://arxiv.org/abs/1003.1643}{{\tt 1003.1643}}.

\bibitem{Alwall:2011uj}
J.~Alwall, M.~Herquet, F.~Maltoni, O.~Mattelaer and T.~Stelzer, \emph{{MadGraph
  5 : Going Beyond}},
  \href{http://dx.doi.org/10.1007/JHEP06(2011)128}{\emph{JHEP} {\bf 1106}
  (2011) 128}, [\href{http://arxiv.org/abs/1106.0522}{{\tt 1106.0522}}].

\bibitem{Frederix:2009yq}
R.~Frederix, S.~Frixione, F.~Maltoni and T.~Stelzer, \emph{{Automation of
  next-to-leading order computations in QCD: The FKS subtraction}},
  \href{http://dx.doi.org/10.1088/1126-6708/2009/10/003}{\emph{JHEP} {\bf 0910}
  (2009) 003}, [\href{http://arxiv.org/abs/0908.4272}{{\tt 0908.4272}}].

\bibitem{Frixione:1995ms}
S.~Frixione, Z.~Kunszt and A.~Signer, \emph{{Three jet cross-sections to
  next-to-leading order}},
  \href{http://dx.doi.org/10.1016/0550-3213(96)00110-1}{\emph{Nucl.Phys.} {\bf
  B467} (1996) 399--442}, [\href{http://arxiv.org/abs/hep-ph/9512328}{{\tt
  hep-ph/9512328}}].

\bibitem{Frixione:1997np}
S.~Frixione, \emph{{A General approach to jet cross-sections in QCD}},
  \href{http://dx.doi.org/10.1016/S0550-3213(97)00574-9}{\emph{Nucl.Phys.} {\bf
  B507} (1997) 295--314}, [\href{http://arxiv.org/abs/hep-ph/9706545}{{\tt
  hep-ph/9706545}}].

\bibitem{Frixione:2002ik}
S.~Frixione and B.~R. Webber, \emph{{Matching NLO QCD computations and parton
  shower simulations}},
  \href{http://dx.doi.org/10.1088/1126-6708/2002/06/029}{\emph{JHEP} {\bf 0206}
  (2002) 029}, [\href{http://arxiv.org/abs/hep-ph/0204244}{{\tt
  hep-ph/0204244}}].

\bibitem{Frederix:2012ps}
R.~Frederix and S.~Frixione, \emph{{Merging meets matching in MC@NLO}},
  \href{http://dx.doi.org/10.1007/JHEP12(2012)061}{\emph{JHEP} {\bf 1212}
  (2012) 061}, [\href{http://arxiv.org/abs/1209.6215}{{\tt 1209.6215}}].

\bibitem{Frederix:2018nkq}
R.~Frederix, S.~Frixione, V.~Hirschi, D.~Pagani, H.~S. Shao and M.~Zaro,
  \emph{{The automation of next-to-leading order electroweak calculations}},
  \href{http://dx.doi.org/10.1007/JHEP07(2018)185}{\emph{JHEP} {\bf 07} (2018)
  185}, [\href{http://arxiv.org/abs/1804.10017}{{\tt 1804.10017}}].

\bibitem{Chen:2017ipx}
C.~Chen, Z.~Cui, G.~Li, Q.~Li, M.~Ruan, L.~Wang et~al., \emph{{$H \rightarrow
  e^+ e^- $ at CEPC: ISR effect with MadGraph}},
  \href{http://arxiv.org/abs/1705.04486}{{\tt 1705.04486}}.

\bibitem{Li:2018qnh}
Q.~Li and Q.-S. Yan, \emph{{Initial State Radiation Simulation with MadGraph}},
   \href{http://arxiv.org/abs/1804.00125}{{\tt 1804.00125}}.

\bibitem{Frixione:2019lga}
S.~Frixione, \emph{{Initial conditions for electron and photon structure and
  fragmentation functions}},
  \href{http://dx.doi.org/10.1007/JHEP11(2019)158}{\emph{JHEP} {\bf 11} (2019)
  158}, [\href{http://arxiv.org/abs/1909.03886}{{\tt 1909.03886}}].

\bibitem{Gribov:1972ri}
V.~N. Gribov and L.~N. Lipatov, \emph{{Deep inelastic e p scattering in
  perturbation theory}}, {\emph{Sov. J. Nucl. Phys.} {\bf 15} (1972) 438--450}.

\bibitem{Lipatov:1974qm}
L.~N. Lipatov, \emph{{The parton model and perturbation theory}}, {\emph{Sov.
  J. Nucl. Phys.} {\bf 20} (1975) 94--102}.

\bibitem{Altarelli:1977zs}
G.~Altarelli and G.~Parisi, \emph{{Asymptotic Freedom in Parton Language}},
  \href{http://dx.doi.org/10.1016/0550-3213(77)90384-4}{\emph{Nucl. Phys.} {\bf
  B126} (1977) 298--318}.

\bibitem{Dokshitzer:1977sg}
Y.~L. Dokshitzer, \emph{{Calculation of the Structure Functions for Deep
  Inelastic Scattering and e+ e- Annihilation by Perturbation Theory in Quantum
  Chromodynamics.}}, {\emph{Sov. Phys. JETP} {\bf 46} (1977) 641--653}.

\bibitem{Skrzypek:1990qs}
M.~Skrzypek and S.~Jadach, \emph{{Exact and approximate solutions for the
  electron nonsinglet structure function in QED}},
  \href{http://dx.doi.org/10.1007/BF01483573}{\emph{Z. Phys.} {\bf C49} (1991)
  577--584}.

\bibitem{Skrzypek:1992vk}
M.~Skrzypek, \emph{{Leading logarithmic calculations of QED corrections at
  LEP}}, {\emph{Acta Phys. Polon.} {\bf B23} (1992) 135--172}.

\bibitem{Cacciari:1992pz}
M.~Cacciari, A.~Deandrea, G.~Montagna and O.~Nicrosini, \emph{{QED structure
  functions: A Systematic approach}},
  \href{http://dx.doi.org/10.1209/0295-5075/17/2/007}{\emph{Europhys. Lett.}
  {\bf 17} (1992) 123--128}.

\bibitem{Bertone:2019hks}
V.~Bertone, M.~Cacciari, S.~Frixione and G.~Stagnitto, \emph{{The partonic
  structure of the electron at the next-to-leading logarithmic accuracy in
  QED}}, \href{http://dx.doi.org/10.1007/JHEP03(2020)135}{\emph{JHEP} {\bf 03}
  (2020) 135}, [\href{http://arxiv.org/abs/1911.12040}{{\tt 1911.12040}}].

\bibitem{Frixione:2021wzh}
S.~Frixione, \emph{{On factorisation schemes for the electron parton
  distribution functions in QED}},  \href{http://arxiv.org/abs/2105.06688}{{\tt
  2105.06688}}.

\bibitem{Maltoni:2002qb}
F.~Maltoni and T.~Stelzer, \emph{{MadEvent: Automatic event generation with
  MadGraph}},
  \href{http://dx.doi.org/10.1088/1126-6708/2003/02/027}{\emph{JHEP} {\bf 0302}
  (2003) 027}, [\href{http://arxiv.org/abs/hep-ph/0208156}{{\tt
  hep-ph/0208156}}].

\bibitem{Chen:1994jt}
P.~Chen, G.~Horton-Smith, T.~Ohgaki, A.~W. Weidemann and K.~Yokoya,
  \emph{{CAIN: Conglomerat d'ABEL et d'interactions nonlineaires}},
  \href{http://dx.doi.org/10.1016/0168-9002(94)01186-9}{\emph{Nucl. Instrum.
  Meth. A} {\bf 355} (1995) 107--110}.

\bibitem{Schulte:1997nga}
D.~Schulte, \emph{{Study of Electromagnetic and Hadronic Background in the
  Interaction Region of the TESLA Collider}}.
\newblock PhD thesis, Hamburg U., 1997.

\bibitem{Schulte:2007zz}
C.~Rimbault, P.~Bambade, O.~Dadoun, G.~Le~Meur, F.~Touze, M.~C. del Alabau
  et~al., \emph{{GUINEA PIG++ : An Upgraded Version of the Linear Collider Beam
  Beam Interaction Simulation Code GUINEA PIG}},
  \href{http://dx.doi.org/10.1109/PAC.2007.4440556}{\emph{Conf. Proc. C} {\bf
  070625} (2007) 2728}.

\bibitem{Ohl:1996fi}
T.~Ohl, \emph{{CIRCE version 1.0: Beam spectra for simulating linear collider
  physics}},
  \href{http://dx.doi.org/10.1016/S0010-4655(96)00167-1}{\emph{Comput. Phys.
  Commun.} {\bf 101} (1997) 269--288},
  [\href{http://arxiv.org/abs/hep-ph/9607454}{{\tt hep-ph/9607454}}].

\bibitem{Sailer:2009zz}
A.~P. Sailer, \emph{{Studies of the measurement of differential luminosity
  using Bhabha events at the International Linear Collider}}.
\newblock PhD thesis, Humboldt U., Berlin, 2009.
\newblock 10.3204/DESY-THESIS-2009-011.

\bibitem{Kilian:2007gr}
W.~Kilian, T.~Ohl and J.~Reuter, \emph{{WHIZARD: Simulating Multi-Particle
  Processes at LHC and ILC}},
  \href{http://dx.doi.org/10.1140/epjc/s10052-011-1742-y}{\emph{Eur. Phys. J.
  C} {\bf 71} (2011) 1742}, [\href{http://arxiv.org/abs/0708.4233}{{\tt
  0708.4233}}].

\bibitem{Yokoya:1989jb}
K.~Yokoya and P.~Chen, \emph{{Electron Energy Spectrum and Maximum Disruption
  Angle Under Multi-Photon Beamstrahlung}}, {\emph{Conf. Proc. C} {\bf 8903201}
  (1989) 1438}.

\end{thebibliography}\endgroup

\end{document}